\documentclass[twocolumn, astrosymb]{aastex7}

\usepackage{microtype} 
\usepackage{color}
\usepackage{amsmath}
\usepackage{gensymb}

\newcommand{\code}[1]{\texttt{#1}}
\newcommand{\phasef}{\Phi(\theta; \lambda)}

\newcommand{\new}[1]{\textcolor{black}{#1}}
\newcommand{\newnew}[1]{\textcolor{black}{#1}}
\newcommand{\obrienzodi}{\text{ZodiSURF}}
\newcommand{\gallat}{b_{\text{gal}}}
\newcommand{\gallon}{l_{\text{gal}}}
\newcommand{\ecllat}{b_{\text{ecl}}}
\newcommand{\ecllon}{l_{\text{ecl}}}
\newcommand{\MJy}{MJy sr$^{-1}$}
\newcommand{\nW}{nW m$^{-2}$ sr$^{-1}$}

\newcommand\cge          {{$\gtrsim$}}
\newcommand\cle          {{$\lesssim$}}
\newcommand\mum          {{\micron}}
\newcommand\eg           {{e.g.},}
\newcommand\ie           {{i.e.},}

\newcommand{\tenexp}[1]{$\times 10^{#1}$}

\defcitealias{Carleton_2022}{SKYSURF-2}
\defcitealias{Windhorst_2022}{SKYSURF-1}
\defcitealias{OBrien_2023}{SKYSURF-4}
\defcitealias{McIntyre_2025}{SKYSURF-6}

\accepted{February 10, 2026}
\submitjournal{ApJ}

\graphicspath{{./}{figures/}}

\begin{document}

\title{SKYSURF-11: A New Zodiacal Light Model Optimized for Optical Wavelengths}


\author[orcid=0000-0003-3351-0878]{Rosalia O'Brien}
\affiliation{School of Earth and Space Exploration, Arizona State University, Tempe, AZ 85287-6004, USA}
\affiliation{UMCP/CRESST2, NASA Goddard Space Flight Center, Greenbelt, MD 20771, USA}
\email[show]{robrien5@asu.edu}

\author[orcid=0000-0001-8403-8548]{Richard G. Arendt}
\affiliation{UMBC/CRESST2, NASA Goddard Space Flight Center, Greenbelt, MD 20771, USA}
\email{richard.g.arendt@nasa.gov}

\author[orcid=0000-0001-8156-6281]{Rogier A.~Windhorst}
\affiliation{School of Earth and Space Exploration, Arizona State University, Tempe, AZ 85287-6004, USA}
\email{Rogier.Windhorst@asu.edu}

\author[orcid=0009-0001-9059-0571]{Tejovrash Acharya}
\affiliation{School of Earth and Space Exploration, Arizona State University, Tempe, AZ 85287-6004, USA}
\email{tachary4@asu.edu}

\author[0000-0002-0882-7702]{Annalisa Calamida}
\affiliation{Space Telescope Science Institute, 3700 San Martin Drive, Baltimore, MD 21218, USA}
\email{calamida@stsci.edu}

\author[orcid=0000-0001-6650-2853]{Timothy Carleton}
\affiliation{School of Earth and Space Exploration, Arizona State University, Tempe, AZ 85287-6004, USA}
\email{tmcarlet@asu.edu}

\author[orcid=0000-0002-2099-639X]{Delondrae Carter}
\affiliation{School of Earth and Space Exploration, Arizona State University, Tempe, AZ 85287-6004, USA}
\email{ddcarte3@asu.edu}

\author[orcid=0000-0003-3329-1337]{Seth H.~Cohen}
\affiliation{School of Earth and Space Exploration, Arizona State University, Tempe, AZ 85287-6004, USA}
\email{seth.cohen@asu.edu}

\author[orcid=0000-0001-8033-1181]{Eli Dwek}
\affiliation{Emeritus, Observational Cosmology Lab, NASA Goddard Space Flight Center, Mail Code 665, Greenbelt, MD 20771, USA}
\affiliation{Research Fellow, Center for Astrophysics — Harvard \& Smithsonian, 60 Garden Street, Cambridge, MA 02138, USA}
\email{elidwek.astro@gmail.gov}

\author[orcid=0000-0003-1625-8009]{Brenda L. Frye}
\affiliation{Department of Astronomy/Steward Observatory, University of Arizona, 933 N. Cherry Avenue, Tucson, AZ 85721, USA}
\email{bfrye@arizona.edu}

\author[orcid=0000-0003-1268-5230]{Rolf A.~Jansen}
\affiliation{School of Earth and Space Exploration, Arizona State University, Tempe, AZ 85287-6004, USA}
\email{rolfjansen.work@gmail.com}

\author[orcid=0000-0003-0214-609X]{Scott J. Kenyon}
\affiliation{Smithsonian Astrophysical Observatory, 60 Garden Street, Cambridge, MA 02138, USA}
\email{kenyon@cfa.harvard.edu}

\author[0000-0002-6610-2048]{Anton M. Koekemoer}
\affiliation{Space Telescope Science Institute, 3700 San Martin Drive, Baltimore, MD 21218, USA}
\email{koekemoer@stsci.edu}

\author[0000-0001-6529-8416]{John MacKenty}
\affiliation{Space Telescope Science Institute, 3700 San Martin Drive, Baltimore, MD 21218, USA}
\email{mackenty@stsci.edu}

\author[0009-0000-4308-6332]{Megan Miller}
\affiliation{School of Earth and Space Exploration, Arizona State University, Tempe, AZ 85287-6004, USA}
\email{mlmill46@asu.edu}

\author[0000-0002-6150-833X]{Rafael Ortiz III}
\affiliation{School of Earth and Space Exploration, Arizona State University, Tempe, AZ 85287-6004, USA}
\email{rortizii@asu.edu}

\author[orcid=0000-0002-9946-5259]{Peter C. B. Smith}
\affiliation{School of Earth and Space Exploration, Arizona State University, Tempe, AZ 85287-6004, USA}
\email{petercbsmith@asu.edu}

\author[orcid=0000-0001-9052-9837]{Scott A. Tompkins}
\affiliation{International Centre for Radio Astronomy Research (ICRAR) and the International Space Centre (ISC), The University of Western Australia, M468, 35 Stirling Highway, Crawley, WA 6009, Australia}
\email{stompkins7192@gmail.com}

\begin{abstract}

We present an improved zodiacal light model, optimized for optical wavelengths, using archival Hubble Space Telescope (HST) imaging from the SKYSURF program. The \cite{Kelsall_1998} model used infrared imaging from the Diffuse Infrared Background Experiment (DIRBE) on board the Cosmic Background Explorer to create a 3D structure of the interplanetary dust cloud. However, this model cannot accurately represent zodiacal light emission outside of DIRBE's nominal wavelength bandpasses, the bluest of which is 1.25 \micron. We present a revision to this model (called \obrienzodi) that incorporates analytical forms of both the scattering phase function and albedo as a function of wavelength, which are empirically determined across optical wavelengths ($0.3-1.6$ \micron) from over 5,000 HST sky surface brightness (sky-SB) measurements. This refined model results in significantly improved predictions of zodiacal light emission at these wavelengths and for Sun angles greater than 80\degree. Fits to HST data show an uncertainty in the model of $\sim$4.5\%. Remarkably, the HST sky-SB measurements show an excess of residual diffuse light (HST Sky -- \obrienzodi\ -- Diffuse Galactic Light) of $0.013\pm 0.006$ \MJy. We suggest that a very dim spherical dust cloud may need to be included in the zodiacal light model, which we present here as a toy model.  

\end{abstract}


\section{Introduction}
\label{sec:zodi_intro}




Zodiacal light is a diffuse glow brightly seen in UV-to-IR observations, caused by sunlight scattering and remitting off interplanetary dust (IPD) particles within our Solar System \citep[\eg][]{Leinert_1998}. The brightness of the IPD is mainly due to scattering at $\lambda \lesssim 3.5$ \micron\ and thermal emission at $\lambda \gtrsim 3.5$ \micron.

IPD grains contributing to zodiacal light typically refers to particles with size range of 1 to 100 \micron\ \citep[\eg][]{Gustafson_1994}, and can vary in shape and composition. They are primarily produced by Jupiter-family comets, with smaller contributions from asteroids and Oort-cloud comets \citep[\eg][]{Dermott_1992, Nesvorny_2010, Rowan-Robinson_2013}.

The IPD cloud structure is shaped by solar radiation, the solar wind, and the Sun’s gravitational and magnetic influence \citep[\eg][]{Burns_1979}. The locations at which grains are released from their parent bodies, and their orbits, also contribute to the overall structure. Individual dust grains are continuously pushed outward by radiation pressure or drawn inward by Poynting–Robertson drag \citep[\eg][]{Mann_2000}, typically on timescales $<10$ Myr \citep{Wyatt_1950}, resulting in the overall brightness of zodiacal light to be remarkably steady on human timescales \citep[\eg][]{Leinert_1982_stability}. 

There are several known components to the IPD cloud \citep[\eg][]{Dermott_1996, Leinert_1998}. The most significant is a large smooth cloud component (with a torus-like shape) that exhibits a slight inclination relative to the ecliptic plane \citep{Leinert_1980}. The density of the smooth cloud decreases with distance from the Sun, following a power-law distribution \citep[\eg][]{Levasseur-Regourd_1996}. We also observe smaller structures, like dust bands due to asteroidal dust \citep{Low_1984, Spiesman_1995}, and a circumsolar ring that contains dust trapped in gravitational resonance with Earth \citep[\eg][]{Reach_2010}. 
The local IPD cloud producing the zodiacal light seen from Earth extends from near the Sun to the asteroid belt at $\sim$3.3 AU, beyond which the zodiacal light becomes negligible \citep[\eg][]{Hanner_1974, Matsumoto_2018, Humes_1980, Stenborg_2024}, \new{though micron-sized IPD particles have been detected as far as 18 AU \citep[\eg][]{Humes_1980, Benn_2017, Jorgensen_2021}.}


At optical wavelengths, the IPD grains that produce zodiacal light are characterized by their albedo and scattering phase function. Albedo refers to the fraction of solar irradiance reflected off the average individual dust particle that scatters at that wavelength. The phase function describes the angular probability of light scattering off dust particles of a given size, effectively acting as a probability density function for scattering angles at that wavelength. Both the albedo and the shape of the scattering phase function, as well as the wavelength dependence of both, can offer insights into the composition of IPD \citep[\eg\ by comparing with properties of asteroid and comets from our Solar System,][]{Yang_2015}.


The scattering of zodiacal light depends on the relative size of a dust particle and wavelength of incoming light, and is often described using Mie scattering theory \citep[\eg][]{Torr_1979}. When a dust grain is larger than the wavelength, the scattered light becomes strongly concentrated in the forward direction, with this effect increasing as the size difference grows. However, Mie theory assumes the dust grains are perfect, homogeneous spheres, which may not reflect their true complexity. 

The brightness and structure of zodiacal light depend on several factors: the observer’s position along Earth’s orbit, the viewing geometry relative to the Sun, and the wavelength of the observations. Over the years, many models have been developed to describe zodiacal light, including phenomenological models \citep[\eg][]{Kelsall_1998, Wright_1998, Hahn_2002, Rowan-Robinson_2013} and physically motivated dynamical models \citep[\eg][]{Jones_1993, Nesvorny_2010, Poppe_2016}. \new{Spectral models often approximate the UV-to-near-IR zodiacal light as a reddened solar spectrum \citep[\eg][]{Leinert_1998, Aldering_2001}, and its absolute brightness has been measured through both direct photometry \citep[\eg][]{Levasseur_1973, Leinert_1981, Hanner_1976, Matsumoto_2018, Krick_2012} and spectroscopy \citep[\eg][]{Matsuura_1995, Matsumoto_1996, Tsumura_2010, Tsumura_2013, Matsuura_2017, Korngut_2022, Hanzawa_2024}.}

The \cite{Kelsall_1998} or \cite{Wright_1998} models (hereafter the Kelsall or Wright models) are widely-used three-dimensional models that make use of daily imaging from NASA’s Cosmic Background Explorer (COBE) Diffuse Infrared Background Experiment (DIRBE). The Kelsall model was the first zodiacal light model to fully leverage data from COBE DIRBE, with the end goal to constrain the cosmic infrared background. This model was defined by temporal variation in the sky brightness, thus producing a three-dimensional, parameterized description of the IPD cloud, spanning wavelengths from 1.25 to 240 \micron.

\citet{Wright_1998} proposed an alternative model, also based on COBE DIRBE data, under the assumption that the faintest 25 \micron\ sky-SB at high ecliptic latitudes originates entirely from zodiacal light. This approach is sensitive to any isotropic component in the zodiacal light signal. However, recent comparisons show that the Wright model likely overestimates the zodiacal light contribution seen by HST \citep[\eg][]{Carleton_2022}.

Together, the Kelsall and Wright models laid the groundwork for modern zodiacal studies, but were designed primarily for infrared wavelengths. These models were optimized for DIRBE's nominal wavelengths only: 1.25, 2.2, 3.5, 4.9, 12, 25, 70, 100, 140, and 240 \micron, and at these wavelengths these models are still considered the state of the art.

\subsection{Zodiacal Light As A Foreground Contaminant}

Zodiacal light is the dominant contributor to the total sky surface brightness (sky-SB) in space-based ultraviolet to infrared observations, accounting for over 90\% of the total signal space telescopes receive \citep[\eg][]{Windhorst_2022}. While this makes it a valuable probe of IPD within our Solar System, it also poses a major challenge: zodiacal light is the brightest foreground that must be subtracted to study the faint universe beyond.

Zodiacal light is just one component of the total sky-SB observed from space, alongside diffuse Galactic light (DGL) and the extragalactic background light (EBL). DGL originates from starlight in our Milky Way that is scattered or re-emitted by interstellar dust. It is brightest toward the Galactic center and is typically isolated by subtracting a zodiacal light model \citep[\eg][]{Brandt_Draine_2012, Ienaka_2013, Arai_2015, Sano_2016, Kawara_2017, Onishi_2018, Chellew_2022}. Accurate zodiacal light modeling is thus a prerequisite for properly characterizing DGL.

The faintest component of the sky-SB, the EBL, is the accumulated light from all stars, galaxies, dust, and active galactic nuclei throughout cosmic history \citep[\eg][]{Franceschini_2008, Hauser_2001, Dominguez_2011, Finke_2010, Driver_2016, Andrews_2018}. It provides critical insights into galaxy formation, black hole growth, and dust evolution over time. However, its direct measurement is notoriously difficult, requiring the careful subtraction of brighter foregrounds like zodiacal light and DGL \citep[\eg][]{Driver_2021}.

EBL predictions from known galaxy counts, known as the integrated galaxy light (IGL), generally fall short of direct EBL measurements by up to an order of magnitude  
\citep[\eg][]{Driver_2016}. The origin of this discrepancy remains a longstanding cosmological problem, with possible sources ranging from missing faint galaxies \citep[\eg][]{Conselice_2016}, the extended outskirts of galaxies \citep[\eg][]{Ashcraft_2023}, intrahalo light \citep[\eg][]{Cooray_2012, Zemcov_2014}, population III stars \citep{Matsumoto_2011}, and instrumental effects \citep{Caddy_2022, McIntyre_2025} to more exotic origins like dark matter or black holes \citep{Maurer_2012, Yue_2013, Kashlinsky_2025}. It also raises the possibility that current zodiacal light models may be underestimating a faint, isotropic component of zodiacal light, thus contributing to an excess known as ``diffuse light'' \citep[\eg][]{Windhorst_2022, Carleton_2022, OBrien_2023, McIntyre_2025}. Even a small modeling error, amounting to $\sim$5\% of the total zodiacal light, could mimic a significant apparent excess in the EBL. Recent New Horizons data, collected beyond 40 AU where zodiacal light is negligible, report a much lower direct measurement of the EBL \citep{Postman_2024}, bringing it into close agreement with IGL estimates and suggesting that earlier discrepancies likely arose from residual zodiacal light contamination.


Aside from measurements of DGL and EBL, understanding our own IPD cloud is useful to interpreting exozodiacal light around other stars. As future missions \citep[\eg\ the Habitable Worlds Observatory,][]{2020decadal} aim to characterize Earth-like planets, understanding and modeling foreground dust in both those systems and our own will be critical, as exoplanet detection and characterization ultimately depend on how well we understand light scattered by Solar System dust \citep[\eg][]{Currie_2025, Hom_2024, Ollmann_2023, Bryden_2023, Guyon_2006, Stark_2014}.

\begin{figure*}
    \centering
    \includegraphics[width=0.9\linewidth]{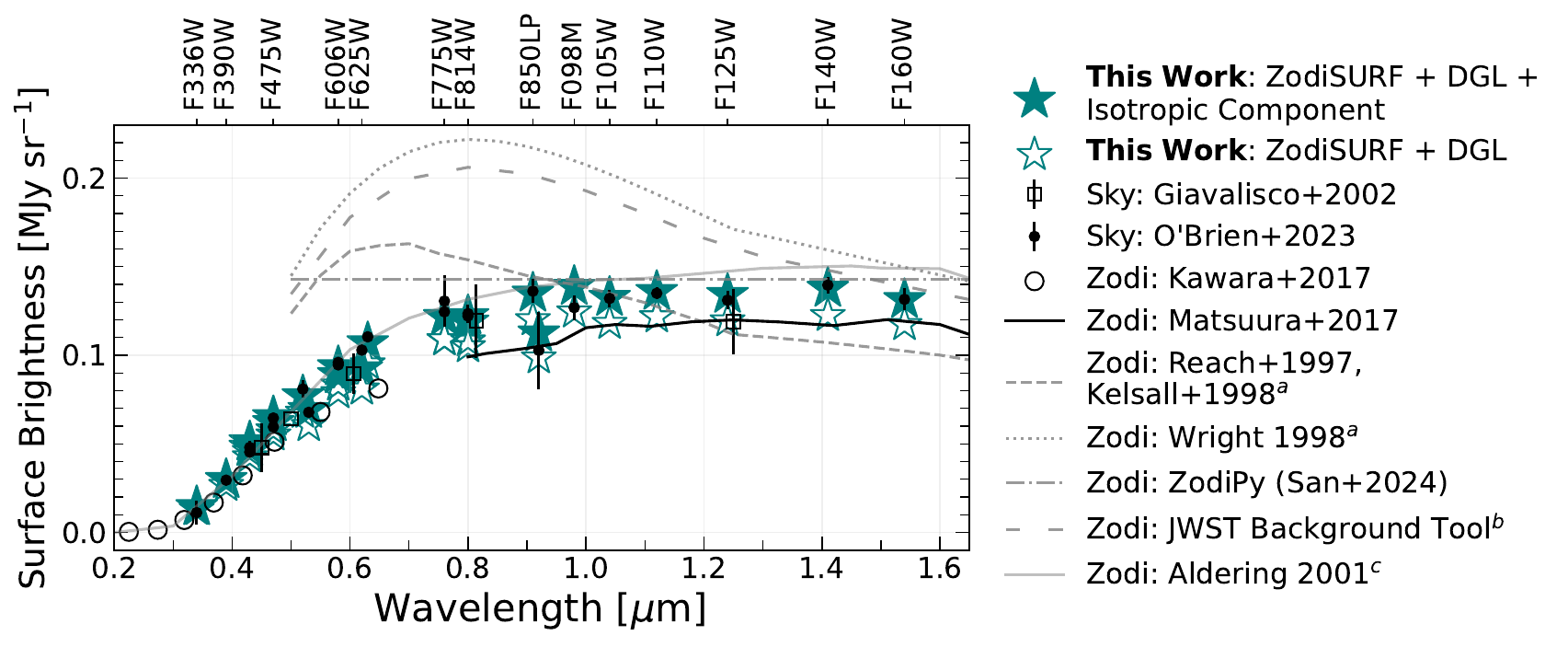}
    \caption{
    \newnew{Comparison of the sky model from this work (teal stars) with HST observed sky-SB (black circles) shows good agreement at 0.3--1.6 \micron.}
    We show other commonly used zodiacal light models \citep{Reach_1997, Kelsall_1998, Wright_1998, Aldering_2001, Rigby_2023, San_2024}, which tend to overpredict at $\lambda \lesssim1.0$ \micron. The SKYSURF HST sky-SB measurements are from \citet{OBrien_2023}, and represent pointings that are within 45\degree\ of the ecliptic poles and are dominated ($>$$90\%$) by zodiacal light. We also show sky-SB measurements from \citet{Giavalisco_2002} as open black squares, and direct measurements of zodiacal light from \cite{Kawara_2017} as open black circles \new{and from \citet{Matsuura_2017} as a solid black line.} \new{The SKYSURF error bars represent the median sky-SB error (as defined in \citeauthor{OBrien_2023} \citeyear{OBrien_2023}) for all pointings represented in a single filter.} The top axis lists the HST filter corresponding to each SKYSURF measurement.
    For these same SKYSURF measurements, we estimate the total sky-SB using the modeling in this work: \obrienzodi$+$DGL (large empty teal stars). We also propose for the addition of an isotropic component to zodiacal light (large filled teal stars).
    We use the IPAC IRSA Background Model\footnote{\url{https://irsa.ipac.caltech.edu/applications/BackgroundModel/}} to retrieve the Kelsall and Wright predictions for those same observations. We use the Cosmoglobe ZodiPy code \citep{San_2024}, the JWST Background Tool\footnote{\url{https://jwst-docs.stsci.edu/jwst-other-tools/jwst-backgrounds-tool}}, and the Gunagala\footnote{\url{https://gunagala.readthedocs.io/en/develop/}} model (implementation of the \cite{Aldering_2001} model) to calculate zodiacal light predictions.
    \newnew{For the Kelsall, Wright, ZodiPy, JWST Background Tool, and Gunagala predictions (shown as dashed and solid grey lines), we plot the true model spectrum at an arbitrary time and location ($\ecllon=0$, $\ecllat=90$, day$=100$) and scale it to match the true median model prediction at 1.25 \micron.}}
    \label{fig:comparing_to_other_models}
\end{figure*}

In addition, accurate zodiacal light modeling is crucial for preparing efficient and reliable sky surveys. Many space-based telescopes, like the Hubble Space Telescope (HST), James Webb Space Telescope (JWST), Euclid space mission, SPHEREx space mission, and Roman Space Telescope, operate in wavelengths where zodiacal light dominates the photon budget, and their science return depends on our ability to model and predict this foreground. Many recent papers adapt the Kelsall or Wright models for use in foreground modeling in the optical and near-IR \citep[\eg][]{Rigby_2023, San_2024, Crill_2025}. Since the Kelsall and Wright models were originally tuned to the DIRBE instrument’s bandpasses (the shortest of which is 1.25 \micron), these models are not easily extrapolated to optical wavelengths. JWST modeling of the sky-SB \citep{Rigby_2023} has shown that these models underperform at $\lambda \lesssim 1$ \micron, highlighting the need for an updated optical-optimized model. To date, no three-dimensional zodiacal light model exists that is optimized specifically for the UV–optical regime. As shown in Figure \ref{fig:comparing_to_other_models}, the Kelsall, Wright, and other models underperform at $\lambda\lesssim1.25$ \micron. Improved zodiacal light models are urgently needed, not only to enable accurate EBL measurements, but also to better understand exozodiacal light, and support the science goals of current and future space observatories.

\subsection{The SKYSURF Project}

The SKYSURF\footnote{\url{http://skysurf.asu.edu}} project \citep[][hereafter referred to as SKYSURF-1 and SKYSURF-2]{Windhorst_2022, Carleton_2022} offers a powerful and data-rich approach to modeling zodiacal light and separating it from other diffuse sky components. SKYSURF is the largest HST archival program to date designed to measure the sky-SB and disentangle its main contributors: the zodiacal light, DGL, and EBL. By leveraging the vast HST archive, spanning multiple instruments, filters, and thousands of independent sky pointings, SKYSURF provides an unprecedented view of how these components vary across the sky at optical wavelengths.

Panchromatic sky-SB measurements were published in \citet[][hereafter referred to as SKYSURF-4]{OBrien_2023}. A key next step is to isolate and quantify the individual contributions of zodiacal light, DGL, and EBL from the observed sky-SB. However, progress has been limited by the lack of a zodiacal light model specifically tuned to optical wavelengths.
In this paper, we present a new zodiacal light model built upon the structure of the widely used Kelsall model framework, but calibrated and optimized for optical observations. We demonstrate that our new model accurately extends coverage to optical wavelengths for the first time. 

In Section \ref{sec:parametric_zodi} of this paper, we summarize our parametric zodiacal light model, which is based on the Kelsall model. In Section \ref{sec:sky_measurement_catalog}, we summarize the HST sky-SB data used to refine the Kelsall model. In Section \ref{sec:dgl_estimates}, we describe the DGL model subtracted from the sky-SB measurements before model fitting. Section \ref{sec:modeling_tech} describes our zodiacal light modeling technique. Section \ref{sec:results} shows our results and improvements on previous models, and Section \ref{sec:discussion} discusses these results in the context of IPD composition and diffuse light residuals present in our model.
Throughout this, we use $\Lambda$CDM cosmology \citep[$H_0$=66.9 km s$^{-1}$ Mpc$^{-1}$, $\Omega_\Lambda$=0.68,][]{Planck2016main} and the AB magnitude scale \citep{OkeGunn1983}.







\section{The Parametric Zodiacal Light Model}\label{sec:parametric_zodi}

The Kelsall model remains a foundational model in foreground analysis, and is used as a baseline for the modeling in this work. COBE’s daily motion sampled zodiacal light at a variety of Sun angles (angles relative to the Sun), enabling constraints on the albedo and scattering phase function.
The Kelsall zodiacal light intensity follows as

\begin{eqnarray}\label{eq:kelsall_zodi}
    Z_{\lambda}(p,t) = \sum_{c} \int n_{c}(X,Y,Z) [
    A_\lambda F^{\odot}_\lambda \Phi_{\lambda}(\theta) + \nonumber \\
    (1-A_\lambda)E_{\lambda}B_{\lambda}(T) K_{\lambda}(T)
    ]ds.
\end{eqnarray}

\noindent The zodiacal light intensity is dependent on time ($t$) and sky position ($p$). The IPD cloud has three main components (denoted with $c$): a smooth cloud, dust bands, and a circumsolar ring. Figure 4 from  \citet{Kelsall_1998} shows a 2D representation of these components. Each has its own density, $n_{c}(X,Y,Z)$, for position $(X, Y, Z)$ within the cloud. The albedo ($A_\lambda$), solar flux ($F^{\odot}_\lambda$), phase function ($\Phi_{\lambda}$), and emissivity ($E_\lambda$) are optimized for each DIRBE nominal wavelength. $B_{\lambda}(T)$ is a Planck blackbody thermal radiance function. $K_{\lambda}(T)$ is a DIRBE color-correction factor from the COBE Diffuse Infrared Background Experiment DIRBE Explanatory Supplement\footnote{\url{https://lambda.gsfc.nasa.gov/data/cobe/dirbe/doc/des_v2_3.pdf}} used to make the model values consistent with the DIRBE calibration. The dust grain temperature $T$ is assumed to vary with distance from the Sun as $T(R) = T_0R^{\delta}$, with a best fit of $T_0 = 286$ K and $\delta = 0.467$. These values are integrated over the line-of-sight ($ds$), where the observer in the DIRBE model is Earth as it orbits around the Sun. 

\begin{figure}
    \centering
    \includegraphics[width=0.7\linewidth]{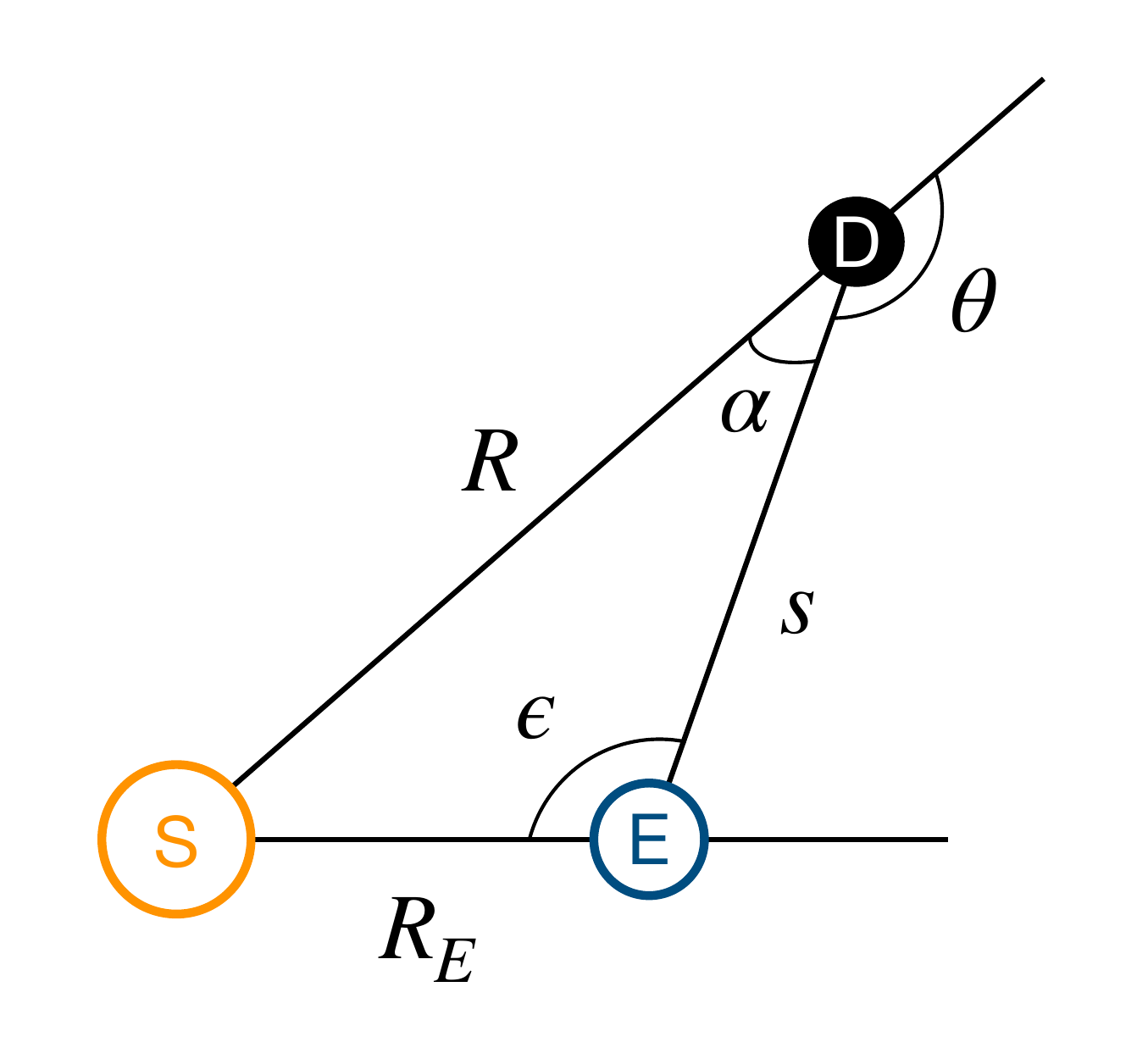}
    \caption{Geometry illustrating the scattering angle ($\theta$), phase angle ($\alpha$), and Sun angle ($\epsilon$) for a dust particle as seen from Earth. The distance from the Sun (S) to the Earth (E) is $R_E$ and from the Sun to the dust particle (D) is $R$. The distance from the Earth to the dust particle is $s$, which represents the model’s line of sight. The scattering angle is given by $\theta = \pi - \alpha$, where $\alpha = \arcsin[(R_E/R)\sin\epsilon]$.}
    \label{fig:kelsall_scatt_cartoon}
\end{figure}

This model optimized various parameters of each component of the dust cloud, including temperature, density, shape, and position. Their optimized values are listed in Tables 1 and 2 of \cite{Kelsall_1998}. We consider their parameters describing the geometry of the IPD to be well optimized, particularly given that the DIRBE long wavelength data constrains the geometry of this emission well. In this work, we focus on optimizing the scattering physics (the scattering phase function and albedo), which is less well constrained and does not extrapolate well to to optical wavelengths. Because DIRBE observations were restricted to Sun angles of 94\degree $\pm$ 30\degree, the model is poorly constrained at large backscattering angles. Backscattering is especially important at very high Sun angles, near 180\degree, where the optical phenomenon known as gegenschein appears \citep[\eg][]{Roosen_1970}. Furthermore, the Kelsall model did not include any component that would appear isotropic (\eg\ a distant, faint spherical shell). Since this model characterized the annual modulation of zodiacal light, any component that appears isotropic across the sky and throughout the year could not have been isolated.

The Kelsall model assumes the albedo is zero for wavelengths greater than 3.5 \micron. It also assumes that the emissivity is zero for wavelengths less than 3.5 \micron. For this work, we will therefore ignore the emissivity, as the thermal emission at 1.6 \micron\ for a 286 K blackbody is negligible. We simplify the Kelsall function (neglecting thermal emission) to be optimized for optical wavelengths, and our model (\obrienzodi) is defined as:

\begin{eqnarray}\label{eq:zodi2025_model}
    \obrienzodi(\lambda, l, b, t) = \nonumber \\
    \sum_{c} \int n_{c}(X,Y,Z) 
    A(\lambda) F^{\odot}(\lambda) \phasef
    ds,
\end{eqnarray}

\noindent where $\theta$ is the scattering angle for the scattering phase function $\phasef$. The ecliptic coordinates ($l,b$) and day of the year ($t$) define the model’s line of sight ($s$). As shown in Figure \ref{fig:kelsall_scatt_cartoon}, this geometry determines the Sun angle ($\epsilon$), phase angle ($\alpha$), and scattering angle ($\theta$) relevant for the dust-scattering model.


\subsection{A New Scattering Phase Function}

The Kelsall model defines the scattering phase function as follows:

\begin{equation}\label{eq:og_phase_func}
    \phasef = N[C_{0,\lambda}+C_{1,\lambda}\theta+e^{C_{2,\lambda}\theta}].
\end{equation}

\noindent The phase function parameters $C_{0,\lambda}$, $C_{1,\lambda}$, and $C_{2,\lambda}$ were specifically fitted to the nominal DIRBE wavelengths (1.25, 2.2, 3.5, 4.9, 12, 25, 70, 100, 140, and 240 \micron). However, this phase function formulation is non-linear. Interpolating between $C_{0,\lambda}$, $C_{1,\lambda}$, and $C_{2,\lambda}$ for various wavelengths does not yield reasonable solutions, which suggest that simple extrapolation of the parameters to optical wavelengths is unreliable.

To provide a better interpolation and extrapolation, we follow \cite{Hong_1985} and replace the Kelsall phase functions with functions that are the weighted sum of 3 Henyey-Greenstein (HG) functions \citep{Henyey_1941}. The benefit of this approach is that it allows the forward and backward scattering strengths to be adjusted separately for each wavelength. As such, this formulation has been widely used to model the scattering of IPD \citep[see, e.g.,][]{Crill_2025}. The new scattering phase function used in this work has the following form:

\begin{equation} \label{eq:new_phase_func}
    \phasef = \sum_{i=1}^{i=3} w_{i}(\lambda) \frac{1-g_{i}(\lambda)^2}{[1+g_{i}(\lambda)^2-2g_{i}(\lambda)\text{cos}(\theta)]^{3/2}}.
\end{equation}

\begin{figure}[t]
    \centering
    \includegraphics[width=0.9\linewidth]{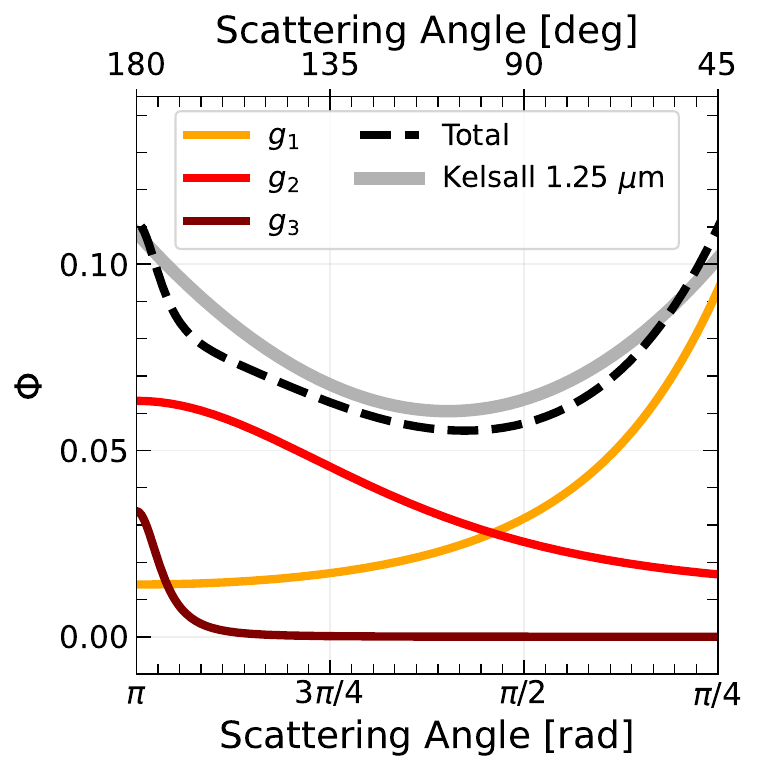}
    \caption{Example of the relative contribution of each $g$ parameter from Equation \ref{eq:new_phase_func}, compared with the Kelsall scattering phase function for $\lambda = 1.25$ \micron. $g_1$ represents the forward scattering component of the scattering phase function, $g_2$ represents the backward scattering component, and $g_3$ represents the gegenschein component. We use values of
    $g_1$ = 0.43, $w_1$ = 0.05,
    $g_2$ = -0.24, $w_2$ = 0.03,
    $g_3$ = -0.87, and $w_3$ = 0.0003. The x-axis is limited to 0.8 radians ($\sim$45\degree), just below the minimum Sun Angle observable by HST.}
    \label{fig:relative_contribution_gparams}
\end{figure}

\noindent In this formulation, we define $i = 1$ to correspond to a forward-scattering component, $i = 2$ to correspond to the backward-scattering component, and $i = 3$ to correspond to the gengenschein component. As an example, Figure \ref{fig:relative_contribution_gparams} illustrates how the relative contributions of each parameter to the total phase function perform at 1.25 \micron. 

\section{Data Overview}\label{sec:sky_measurement_catalog}

To optimize the scattering phase function and albedo, we use sky-SB measurements from the SKYSURF project (\citetalias{OBrien_2023}; \citeauthor{McIntyre_2025} \citeyear{McIntyre_2025}, hereafter referred to as SKYSURF-6). SKYSURF developed an algorithm that measures the object-free sky-SB in HST images numerically to better than 1\% precision, with total calibration uncertainties below 4\%. The public SKYSURF release contains over 150,000 sky-SB measurements spanning the full HST wavelength range, from 0.2 to 1.6 \micron. These measurements cover a wide range of Sun angles (from 80\degree\ to 180\degree), allowing HST to probe a broad range of scattering angles. This broad wavelength and Sun angle coverage is crucial for constraining how the zodiacal light phase function and albedo evolve with wavelength, especially into the ultraviolet, where few other measurements exist. Each measurement includes an uncertainty that accounts for algorithmic errors, flat-fielding, photometric zeropoints, WFC3/IR nonlinearity, bias, dark current, post-flash, and thermal dark contributions \citepalias[for the latter, see][]{McIntyre_2025}.

To ensure data quality, SKYSURF implements a detailed flagging system to identify reliable measurements for studying zodiacal light and EBL. Following \citetalias{OBrien_2023}, we exclude all sky-SB values flagged as unreliable. Based on the discussion in \citetalias[][]{Windhorst_2022} and \citetalias[][]{Carleton_2022}, images are flagged as having potentially too high sky-SB levels to accurately trace the foreground light if one or more of the following conditions applied (in no particular order):
\begin{itemize}
    \item More than 30\% of the image is contaminated by large, discrete galaxies or stars, as determined by the SKYSURF sky-SB algorithm.
    \item The sky-SB rms is higher than 2.5$\times$ the expected rms, based on Gaussian and Poisson noise.
    \item The image was flagged during manual inspection.
    \item The image is taken within 20\degree\ of the Galactic plane.
    \item The image is taken with an Earth limb angle of less than 40\degree.
    \item The image is taken with a Sun altitude greater than 0\degree.
    \item The image is taken with a Sun angle less than 80\degree.
    \item The image is taken with a Moon angle less than 50\degree.
    \item The image is taken close to large (in terms of spatial sky area), nearby galaxies.
    \item The image is significantly affected by persistence.
\end{itemize}

Because many HST observations revisit the same sky positions, either to build depth or monitor variability, we avoid biasing the model toward oversampled regions by restricting our sample to ``independent pointings''. We define a pointing as independent if it is taken at least 10 arcminutes and 2 days apart from any other observation. However, revisiting a region does improve sky-SB measurement reliability. So, for dependent pointings (those close in sky position and time), we group them and compute a single representative measurement. For each group, we take the median values of sky position (ecliptic and Galactic coordinates), day-of-year, sky-SB, and Sun angle. The group’s uncertainty is the median individual uncertainty divided by the square root of the number of images in the group. This reduces statistical noise without over-weighting heavily observed fields.

We apply thermal dark corrections (cased by thermal emission from HST's instruments) using the refined estimates from \citetalias{McIntyre_2025}, which are based on HST’s pick-off arm temperature and empirical models of the primary and secondary mirror thermal emission levels. These thermal dark values are listed in Table \ref{tab:thermal_dark} and are subtracted from the original SKYSURF sky-SB values.

\begin{table}[h]
    \centering
    \caption{Thermal dark corrections used in this work \citepalias[from][]{McIntyre_2025}.}
    \begin{tabular}{ccc}
        Filter & Thermal Dark (e/s/px) & Thermal Dark (MJy/sr) \\
        \hline
        \hline
        F098M  & 0.0044 & 0.0023 \\
        F105W  & 0.0044 & 0.0013 \\
        F110W  & 0.0047 & 0.0008 \\
        F125W  & 0.0047 & 0.0014 \\
        F140W  & 0.0217 & 0.0054 \\
        F160W  & 0.0820 & 0.0327 \\
        \hline
    \end{tabular}
    \label{tab:thermal_dark}
\end{table}

The final number of independent, quality-controlled sky-SB measurements used for this zodiacal light analysis is a few dozen to many hundreds, depending on the filter, and is summarized in Table \ref{tab:sky_catalog_table}.

\begin{table*}
    \centering
    \caption{HST filters used to calibrate the zodiacal light model in this work. We list the HST camera, the filter name, and the corresponding pivot wavelength. The fourth column shows the number of SKYSURF sky-SB measurements in that filter used for model fitting. The fifth columns shows the DGL scale factor ($C_{\mathrm{DGL},\lambda}$ in Equation \ref{eq:my_dgl}) used to compute the final DGL for each pointing. The sixth column shows the median DGL level for each filter. The final column shows the solar irradiance level used for each filter.}
    \begin{tabular}{ccccccc}
Camera & Filter & $\lambda$ [\micron] & Number of Images & $C_{\mathrm{DGL},\lambda}$ & DGL [\MJy] & Solar Irradiance [\MJy] \\
\hline
\hline
WFC3/UVIS & F336W & 0.34 & 107 & 0.34 & 0.0007 & 3.341\tenexp{7} \\
WFC3/UVIS & F390W & 0.39 & 84 & 0.46 & 0.0009 & 7.073\tenexp{7} \\
WFC3/UVIS & F438W & 0.43 & 25 & 0.61 & 0.0012 & 1.145\tenexp{8} \\
WFC3/UVIS & F475X & 0.48 & 14 & 0.78 & 0.0015 & 1.488\tenexp{8} \\
WFC3/UVIS & F475W & 0.47 & 57 & 0.74 & 0.0014 & 1.455\tenexp{8} \\
WFC3/UVIS & F555W & 0.52 & 22 & 0.91 & 0.0018 & 1.772\tenexp{8} \\
WFC3/UVIS & F606W & 0.58 & 253 & 1.07 & 0.0021 & 1.995\tenexp{8} \\
WFC3/UVIS & F625W & 0.62 & 15 & 1.18 & 0.0023 & 2.154\tenexp{8} \\
WFC3/UVIS & F775W & 0.76 & 13 & 1.31 & 0.0026 & 2.377\tenexp{8} \\
WFC3/UVIS & F850LP & 0.92 & 19 & 1.02 & 0.0020 & 2.410\tenexp{8} \\
WFC3/UVIS & F814W & 0.80 & 289 & 1.25 & 0.0024 & 2.385\tenexp{8} \\
ACS/WFC & F435W & 0.43 & 225 & 0.60 & 0.0012 & 1.118\tenexp{8} \\
ACS/WFC & F475W & 0.47 & 190 & 0.73 & 0.0014 & 1.432\tenexp{8} \\
ACS/WFC & F555W & 0.53 & 42 & 0.93 & 0.0018 & 1.814\tenexp{8} \\
ACS/WFC & F606W & 0.58 & 726 & 1.08 & 0.0021 & 2.008\tenexp{8} \\
ACS/WFC & F625W & 0.63 & 72 & 1.20 & 0.0023 & 2.177\tenexp{8} \\
ACS/WFC & F775W & 0.76 & 255 & 1.31 & 0.0026 & 2.378\tenexp{8} \\
ACS/WFC & F814W & 0.80 & 1258 & 1.25 & 0.0024 & 2.386\tenexp{8} \\
ACS/WFC & F850LP & 0.91 & 543 & 1.06 & 0.0021 & 2.406\tenexp{8} \\
WFC3/IR & F098M & 0.98 & 71 & 0.87 & 0.0017 & 2.419\tenexp{8} \\
WFC3/IR & F105W & 1.04 & 353 & 0.82 & 0.0016 & 2.386\tenexp{8} \\
WFC3/IR & F110W & 1.12 & 214 & 0.77 & 0.0015 & 2.347\tenexp{8} \\
WFC3/IR & F125W & 1.24 & 357 & 0.71 & 0.0014 & 2.304\tenexp{8} \\
WFC3/IR & F140W & 1.41 & 336 & 0.65 & 0.0013 & 2.228\tenexp{8} \\
WFC3/IR & F160W & 1.54 & 820 & 0.59 & 0.0012 & 2.125\tenexp{8} \\
\hline
    \end{tabular}
    \label{tab:sky_catalog_table}
\end{table*}

\section{Diffuse Galactic Light Estimates}\label{sec:dgl_estimates}

DGL can be highly uncertain at optical wavelengths due to uncertainties in its scattering properties. Measuring DGL directly can be difficult since zodiacal light is almost always a foreground contaminant. Although we expect DGL estimates to be less than $\sim$$0.003$ \MJy\ for most SKYSURF images used in this work \citepalias[see ][]{Carleton_2022}, we still aim for accurate estimates of it.

To obtain an optical DGL measurement for each HST pointing, we follow the methods presented in \cite{Postman_2024}. This work obtained measurements of the cosmic optical background using the Long-Range Reconnaissance Imager (LORRI) onboard NASA’s New Horizons spacecraft. At nearly 57 AU from the sun, there is virtually no zodiacal light. This means that any empirical methods to estimate DGL with New Horizons will be independent from highly uncertain zodiacal light models. It is standard to correlate the optical DGL emission with the 100 \micron\ sky intensity \citep[\eg][]{Arai_2015, Brandt_Draine_2012, Guhathakurta_1989, Ienaka_2013, Kawara_2017, Onishi_2018, Symons_2023, Witt_2008, Zagury_1999}. However, in developing an independent DGL estimator, \cite{Postman_2024} found that 350 \micron\ and 550 \micron\ were better indicators. They suggest that this is due to variations in dust temperature: the 100 \micron\ band falls on the shorter-wavelength side of the peak for a 20 K blackbody dust spectrum, making it more sensitive to temperature changes compared to the 350 \micron\ and 550 \micron\ bands.

With 24 fields, they empirically related Planck 350 \micron\ and 550 \micron\ imaging and their measured DGL, where their measured DGL is assumed to be the total sky-SB they measure with all other known components (faint stars, stray light from outside the field of view, EBL) subtracted. They find the following relationship for the LORRI bandpass (centered at ${\sim}0.6$ $\micron$):

\begin{multline}\label{eq:postman_dgl}
    \text{DGL}_{\text{Postman}}(\gallon,\gallat)\ [\text{nW m$^{-2}$ sr$^{-1}$}] = \\
    g(\gallat) [
    48.01\ I_{550}(l_{\text{gal}},b_{\text{gal}}) + \\
    0.96 \left( \frac{I_{550}(\gallon,\gallat)}{I_{350}(\gallon,\gallat)}-3.66\right)
    ],
\end{multline}

\noindent where

\begin{equation}
    g(\gallat) = \frac{1-0.67\sqrt{\sin(|\gallat|)}}{0.376}.
\end{equation}


\noindent The DGL intensity is dependent directly on Galactic coordinates ($\gallon,\gallat$). $I$ is the CIB-subtracted FIR intensity: $I(\lambda)$ -- $\text{CIB}_{\text{Anisotropies}}(\lambda)$ -- 
$\text{CIB}_{\text{Monopole}}$. The Planck High Frequency Instrument (HFI) maps \citep{Planck_2020} provide the FIR intensities at 350 \micron\ and 550 \micron. The CIB$_{\text{Anisotropies}}(\lambda)$ maps are provided by \cite{Planck_2016}, and use the generalized needlet internal linear combination (GNILC; see \citeauthor{Remazeilles_2011} \citeyear{Remazeilles_2011}) method to separate CIB anisotropies from thermal dust emission in the HFI maps. These are essentially a field-dependent correction to the CIB monopole, and remain necessary when working with HST's small field-of-view. CIB$_{\text{Monopole}}$ is the CIB monopole as measured by recalibrating Planck HFI maps, separating the Galactic emission using the HI column density, and determining the CIB monopole by extrapolating the HI density to zero \citep{Odegard_2019}, resulting in 0.576 \MJy\ for 350 \micron\ and 0.371 \MJy\ at 550 \micron.

The function \( g(\gallat) \) accounts for changes in scattering effects based on galactic latitude (\( \gallat \)) during the conversion of thermal emission intensity into optical intensity, where scattering is more significant \citep[\eg][]{Jura_1979,Zemcov_2017}. Since the scattering properties change as a function of wavelength, we rewrite \( g(\gallat) \) following \cite{Zemcov_2017}:

\begin{equation}
    g(\lambda, \gallat) = \frac{1-1.1\times f_{\lambda} \times \sqrt{\sin(|\gallat|)}}{1-1.1\times f_{\lambda} \times \sqrt{\sin(60\degree)}}.
\end{equation} \label{eq:final_dgl_asym}

\noindent The scattering asymmetry factor ($f_{\lambda}$) represents the degree of forward scattering. It is defined as $f_{\lambda} \equiv \langle \cos \theta \rangle$, where $\theta$ denotes the scattering angle from the forward direction \citep[\eg][]{Sano_2016}, meaning that $f_{\lambda} = 0$ represents isotropic scattering and $f_{\lambda} = 1$ represents completely forward scattering. This is equivalent to $g$ in the HG function (Equation \ref{eq:new_phase_func}), but we use $f_{\lambda}$ to differentiate DGL scattering from that of zodiacal light. We normalize $g(\gallat)$ at a galactic latitude of 60\degree, so that $g(\gallat)$ only accounts for the relative changes in scattering properties.

\cite{Weingartner_2001} evaluate $f_{\lambda}$ (they label it as $g$) for several size distributions of carbonaceous and silicate grain populations in different regions of the Milky Way, LMC, and SMC. Their Figure 15 shows how $f_{\lambda}$ varies as a function of wavelength. We adopt the results in Figure 15 of \cite{Weingartner_2001} for $f_{\lambda}$. 


Equation \ref{eq:postman_dgl} is specifically designed for the LORRI instrument. To apply this approach to the HST bandpasses, we first fold both the LORRI bandpass and each HST bandpass with a reference DGL spectrum. By comparing the results, we can scale Equation \ref{eq:postman_dgl} based on how the LORRI instrument relates to the HST bandpasses. To calculate the scaling factor, we use the reference DGL spectrum from \cite{Brandt_Draine_2012}. This spectrum is created using sky spectra from the Sloan Digital Sky Survey (SDSS), and therefore only extends from $\sim$$0.4$ to 0.9 $\micron$. For longer wavelengths, we logarithmically interpolate between the \cite{Brandt_Draine_2012} spectrum and the AKARI (the JAXA infrared astronomical satellite) spectrum from \cite{Tsumura_2013}, which starts at $\sim$$1.8$ \micron. We normalize the spectrum so the LORRI result is defined to be equal to 1.0. The DGL measurement for each HST filter is scaled accordingly by a factor of $C_{\mathrm{DGL},\lambda}$.

Our final DGL estimator is therefore
\begin{multline}\label{eq:my_dgl}
    \text{DGL}(\lambda,\gallon,\gallat)\ [\text{nW m$^{-2}$ sr$^{-1}$}] = \\
    g(\lambda, \gallat) [
    48.01\ I_{550}(l_{\text{gal}},b_{\text{gal}}) + \\
    0.96 \left( \frac{I_{550}(\gallon,\gallat)}{I_{350}(\gallon,\gallat)}-3.66\right)
    ] \times C_{\mathrm{DGL},\lambda}.
\end{multline}

\noindent The values for $C_{\mathrm{DGL},\lambda}$ are listed in Table \ref{tab:sky_catalog_table}. $I_{350}$ and $I_{550}$ are estimated using the Planck HFI \code{HEALPix}\footnote{https://healpy.readthedocs.io/en/latest/} maps. We isolate a circular area with a diameter close to that of HST's field of view (202\arcsec\ for ACS/WFC, 162\arcsec\ for WFC3/UVIS, 130\arcsec\ for WFC3/IR) in each \code{healpy} map at the location of the SKYSURF image.

Uncertainties in DGL are taken from \citet[][]{Postman_2024} using the average from their Table 4, which is 1.03 nW m\(^{-2}\) sr\(^{-1}\). This uncertainty represents the rms obtained in their empirical DGL calculation, and was calculated using 10,000 Monte Carlo realizations. We estimate the uncertainty in the \cite{Brandt_Draine_2012} spectrum using Figure 3 from their paper: $\sigma (\lambda I_\lambda / \nu I_\nu (100 \micron)) \simeq 0.01$. This uncertainty represents ${\sim}6\%$ of the peak of the DGL spectrum, or $\lambda I_\lambda / \nu I_\nu (100 \micron) \simeq 0.17$. Therefore, we adopt an uncertainty in the DGL spectrum to be 6\%.

\begin{figure*}
    \centering
    \includegraphics[width=0.8\linewidth]{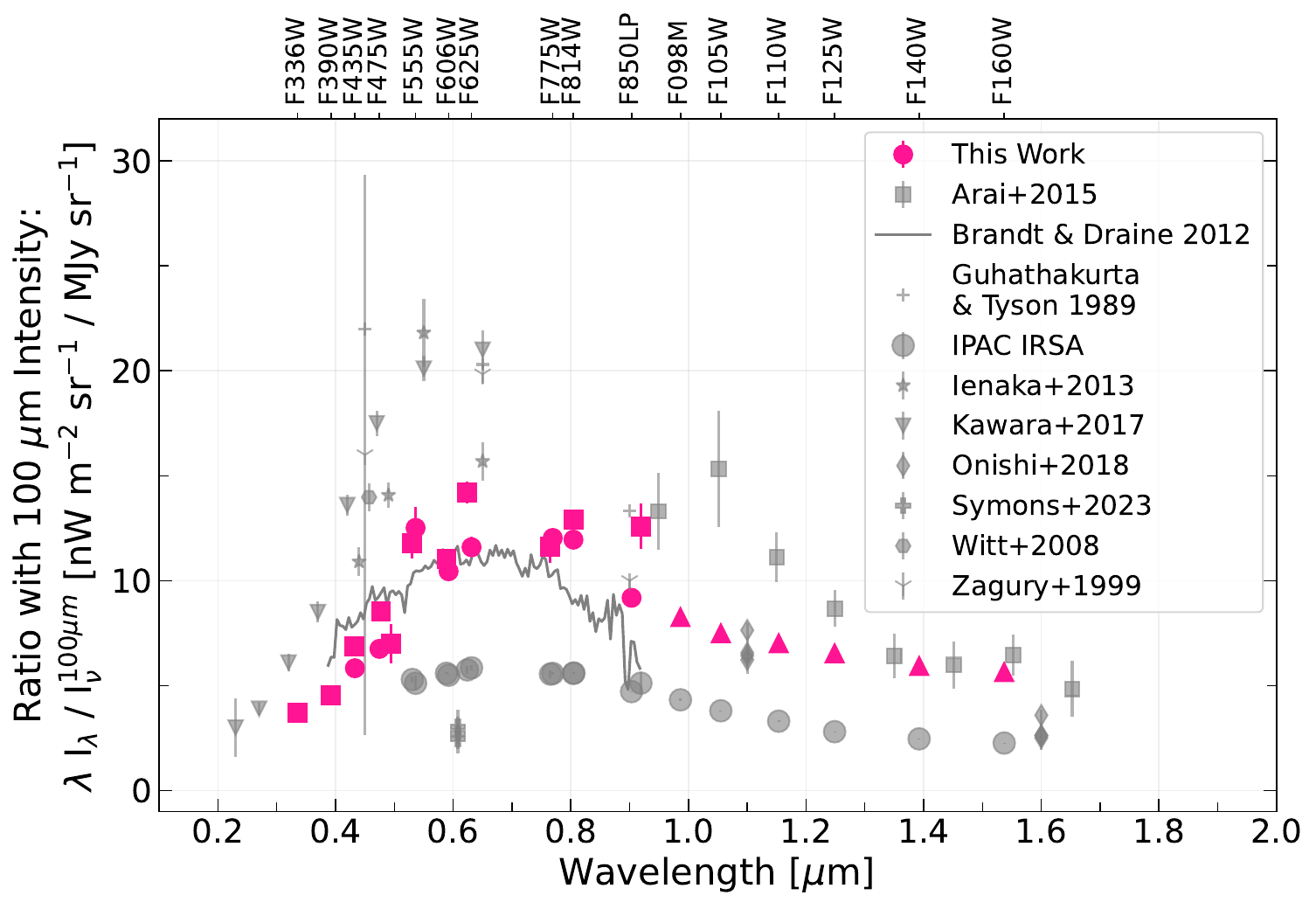}
    \caption{Ratio of DGL intensity (nW m\(^{-2}\) sr\(^{-1}\)) to 100 \micron\ intensity (MJy sr\(^{-1}\)). The 100 \micron\ intensity for both the IPAC IRSA Background Model and this study's comparisons come from the IRIS+SFD maps released in \href{https://irsa.ipac.caltech.edu/data/Planck/release\_2/external-data/}{Planck Data Release 2}. DGL estimates from this work are shown as large pink symbols, where HST's three main cameras are distinguished by different symbols: squares for WFC3/UVIS, circles for ACS/WFC, and triangles for WFC3/IR. The error bars are calculated as the standard deviation of the measurements divided by \(\sqrt{N}\), where \(N\) is the number of independent pointings. The grey symbols represent DGL estimates from various other studies \citep{Guhathakurta_1989, Zagury_1999, Witt_2008, Brandt_Draine_2012, Ienaka_2013, Arai_2015, Kawara_2017, Onishi_2018, Symons_2023}, where it is typically calibrated with 100 \micron\ maps. The SKYSURF project originally used the IPAC IRSA Background Model to estimate DGL, shown as large grey circles. The top axis lists the HST filter corresponding to each DGL measurement from this work.}
    \label{fig:compare_dgl_other_works}
\end{figure*}

Our total DGL is a combination of two independent estimators: the \cite{Postman_2024} measurement of DGL multiplied by a scale factor $C_{\mathrm{DGL},\lambda}$. The uncertainties for these two estimators are described in the previous paragraph. Therefore, the final uncertainty in DGL is:


\begin{multline}
    \sigma_{\text{DGL}}\ [\text{nW m\(^{-2}\) sr\(^{-1}\)}] = \nonumber \\
    |\text{DGL}| \sqrt{\left( \frac{1.03}{\text{DGL}_{\text{Postman}}} \right)^2 + \left( \frac{0.06 \times \text{DGL}}{C_{\mathrm{DGL},\lambda}} \right)^2 }.
\end{multline}

\noindent Final median DGL estimates used for each HST filter are listed in Table \ref{tab:sky_catalog_table}. To minimize the impact of the DGL on our residual diffuse sky signal, we ignore all images taken within 20$\degree$ of the Galactic plane (see Section \ref{sec:sky_measurement_catalog}). The average DGL uncertainty is ${\sim}0.0003$ \MJy. 


We compare our DGL estimates with other work in Figure \ref{fig:compare_dgl_other_works}. Since most studies are correlated with 100 \micron\ intensity, we also compare our measured DGL intensity to the 100 \micron\ maps. For this, we use a combination of the Improved Reprocessing of the IRAS Survey (IRIS) maps \citep{Miville_2005} and the Schlegel, Finkbeiner, and Davis \citep[][SFD]{Schlegel_1998} map released in the Planck Data Release 2 \citep{Planck_2014_thermal_dust}. Specifically, we use the 100 \micron\ combined IRIS+SDF map with no point sources. In general, the methods described in this work agree well with other studies. The scatter in DGL levels shown in Figure 4 across different studies is likely due to variations in the sky regions being observed, as different areas have different DGL contributions and 100-micron calibrations. This is especially true for our data points, as each represents a different combination of regions of the sky. Since we did not use 100 \micron\ maps for calibration, there will naturally be some scatter compared to the 100 \micron\ map.

This DGL model can be improved further, but since this is not the focus of this work, we leave this to future work. Appendix \ref{app:dgl_improvements} has further discussion on the current limitations of our DGL model and suggestions for further improvement.

\section{Zodiacal Light Modeling Technique}\label{sec:modeling_tech}

Our goal is to modify the Kelsall model so that it is accurate at any wavelength between $0.3-1.6$ \micron. Within the Kelsall model itself, this involves updating the solar irradiance spectrum ($F^{\odot}_\lambda$), the scattering phase function ($\phasef$), and the albedo ($A(\lambda)$). The scattering phase function and the albedo will be optimized with HST data, while the solar irradiance spectrum is pulled from recent literature, as described below. 

\subsection{Updating the Solar Irradiance Spectrum}


The current Kelsall model has individual solar irradiance measurements for every DIRBE band. For this model to apply to HST data, we need to provide solar irradiance values for all HST bands. We use the Hybrid Solar Reference Spectrum (HSRS) from \cite{Coddington_2021}, which present a new solar irradiance reference spectrum representative of solar minimum conditions. They use a combination of measurements from NASA’s Total and Spectral Solar Irradiance Sensor \citep[TSIS-1,][]{Richard_2024} and the CubeSat Compact SIM. The TSIS-1 HSRS spans $202-2730$ nm at $0.01$ to $\sim$$0.001$ nm spectral resolution. Their uncertainties are 0.3\% at wavelengths between 460 and 2365 nm, and $\sim$1.3\% at wavelengths outside that range.
Each HST filter bandpass is folded with the TSIS-1 HSRS spectral irradiance for the modeling in this work. The solar irradiance values used in this work are listed in Table \ref{tab:sky_catalog_table}. 

Another recent solar irradiance spectrum is that from \cite{Meftah_2018}. They provide a reference solar irradiance spectrum from the SOLSPEC instrument on the International Space Station. The instrument accurately measured solar spectral irradiance from 0.1 to 3 \micron, resulting in a high-resolution solar spectrum with an average absolute uncertainty of 1.26\%. The spectrum has a varying spectral resolution between 0.6 and 9.5 nm. When comparing this reference spectrum to the TSIS-1 HSRS reference spectrum, offsets are within 4\% in any given filter. Therefore, the uncertainty in the solar irradiance for a particular bandpass is estimated to be typically $\sim$3\%.

\subsection{Fitting a New Phase Function and Albedo}\label{sec:fitting}

The most critical component of the model update is the optimization of the scattering phase function and dust albedo across the wavelength range $0.3$–$1.6$ \micron. Our goal is to empirically derive analytical scattering phase function and albedo functions that best reproduce panchromatic HST sky-SB measurements. To do this, we use sky-SB measurements from \citetalias{OBrien_2023} (as described in Section~\ref{sec:sky_measurement_catalog}) which include contributions from zodiacal light, DGL, and EBL. Since SKYSURF is primarily designed to constrain the EBL and any residual diffuse light, we treat the residual diffuse light as a free parameter in our modeling. The DGL component is subtracted using the methods described in Section~\ref{sec:dgl_estimates}, and we assume its spatial and spectral properties are well-characterized.

We define the observed zodiacal light intensity as:
\begin{multline}\label{eq:observed_zodi}
Z(\lambda,l,b, t) = S(\lambda,l,b, t) \\
- DGL(\lambda,\gallon,\gallat) - C(\lambda),
\end{multline}
where $\lambda$ is the wavelength, $(l,b)$ are ecliptic coordinates, $t$ is the day of year, $S$ is the measured sky-SB from \citetalias{OBrien_2023}, and $C(\lambda)$ represents any remaining isotropic EBL and diffuse light component. Although the EBL contributes to $C(\lambda)$, most of it is removed during measurement of the HST sky-SB, since most discrete stars and galaxies are detected (to AB\cle 27 mag) and masked out before the HST sky-SB measurements take place. The EBL from undetected galaxies with AB\cge 27 mag that are left in sky-SB measurements is estimated to be only $\sim$$0.56$ nW m$^{-2}$ sr$^{-1}$ ($<$$0.0002$ MJy sr$^{-1}$; \citetalias{Carleton_2022}).

The uncertainty in $Z$ is computed as:
\begin{equation}
\sigma_Z = \sqrt{\sigma_S^2 + \sigma_{\text{DGL}}^2},
\end{equation}
where $\sigma_S$ and $\sigma_{\text{DGL}}$ are the uncertainties in the sky-SB and DGL estimates, respectively. The uncertainty in the sky-SB measurement dominates, resulting in an average uncertainty in (Sky-SB -- DGL) to be $\sim$0.005 \MJy. We do not include uncertainties for $C(\lambda)$, as it is treated as a free parameter. Our fitting pipeline proceeds in three steps, which are outlined in the next three sub-sections. 

 
\subsubsection{Step 1: Joint Fit of Albedo and Phase Function}

We jointly fit the albedo and the six phase function parameters ($g_1$, $g_2$, $g_3$, $w_1$, $w_2$, $w_3$) defined in Equation~\ref{eq:new_phase_func}, using the model defined in Equation~\ref{eq:zodi2025_model}. In this step, we fit for each HST filter independently. We use the Markov Chain Monte Carlo (MCMC) sampler  \texttt{emcee} \citep{Foreman-Mackey_2013} to maximize the log-likelihood
\begin{equation}\label{eq:likelihood}
\log \mathcal{L}_j = -\frac{1}{2} \sum_{i=1}^{n_j} 
\left[
\frac{(Z_i - \obrienzodi_i)^2}{\sigma_{Z,i}^2} + \ln(2\pi \sigma_{Z,i}^2)
\right],
\end{equation}
for $n_j$ total HST pointings in filter $j$. Our priors on the free parameters are shown in Table \ref{tab:priors}. We use these priors for both steps 1 \& 2 of our fitting pipeline.

\begin{table}[h]
    \centering
    \caption{Uniform priors used for zodiacal light model parameter fitting.}
    \begin{tabular}{cc}
        Parameter & Prior \\
        \hline
        \hline
        Isotropic Scale $C(\lambda)$ & [0,0.3]\\
        Albedo & [-$\infty$, $\infty$]\\
        $g_1$ & [0,1]\\
        $g_2$ & [-1,0]\\
        $g_3$ & [-1,-0.6]\\
        $w_1$ & [0,1]\\
        $w_2$ & [0,1]\\
        $w_3$ & [0,1]\\
        \hline
    \end{tabular}
    \label{tab:priors}
\end{table}

Since the phase function acts like a probability function, we require it to integrate to 1.0 over $4\pi$ steradians:
\begin{equation}
\int_0^{2\pi} \int_0^{\pi} \phasef \sin(\theta) d\theta d\phi = 1.
\end{equation}
Because the HG function lacks a closed-form integral, we compute this numerically using \texttt{scipy.integrate.simpson} \citep{SciPy_2020}. We apply a Gaussian prior with this condition to our \texttt{emcee} routine, centered at 1.0 with a standard deviation of 0.001.

For each HST filter, we use 25 MCMC walkers and run for 27,000 iterations. The first 15,000 steps are discarded as burn-in, althought most filters converge within 1,000 iterations (verified by visual inspection). No other priors are applied. This step yields posterior distributions for the eight parameters (albedo, $g_i$, $w_i$, and $C(\lambda)$) per filter. 

\begin{figure}
    \centering
    \includegraphics[width=\linewidth]{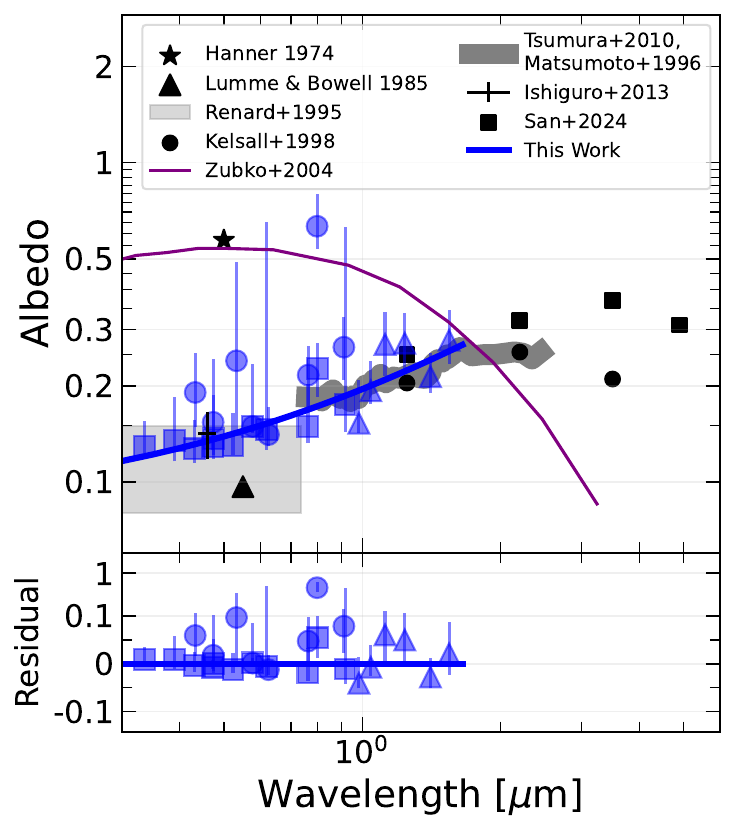}
    \caption{Albedo measurements from this work (solid blue line) compared to values from previous studies. Blue symbols with error bars show the best-fit albedos for each HST filter. The error bars represent the 68.27th-percentile distribution from the posterior. Different symbol shapes represent different HST cameras: squares represent WFC3/UVIS, circles represent ACS/WFC, and triangles represent WFC3/IR. For reference, the solid black circles show the original albedo model from \citet{Kelsall_1998}. Black symbols show albedo measurements from other studies: \citet{Hanner_1974} (Helios A and Pioneer 10), \cite{Lumme_1985} (using polarimetric data), \cite{Renard_1995} (using nodes of lesser uncertainty method), \citet{Ishiguro_2013} (from the WIZARD instrument), and \citet{San_2024} (a reanalysis of COBE/DIRBE data). The first two report geometric albedos, which are converted to single-particle albedos for this comparison (conversion shown in Appendix \ref{app:geometric_albedo}). \new{The grey curve represents the relative zodiacal light reflectance spectrum from CIBER and IRTS \citep{Tsumura_2010, Matsumoto_1996}, scaled to the Kelsall albedo at 2.2 \micron.} The purple curve shows the modeled albedo of interstellar dust from \citet{Zubko_2004}. In the bottom panel, we show the best fit albedo for each individual HST filter (blue symbols) with the linear fit (solid blue line) subtracted. The y-axis of the bottom panel is scaled linearly from -0.1 to 0.1, outside of which it is scaled logarithmically. }
    \label{fig:comparing_albedos}
\end{figure}

The phase function and albedo are correlated, so we first constrain the albedo in step 1 of our fitting process. We then fix this constrained albedo when focusing on the phase function in step 2. From step 1, we extract the best-fit albedos and fit a linear trend across wavelength (Figure \ref{fig:comparing_albedos}). The reduced chi-squared for this linear relation is $\chi^2_\nu = 1.06$, and this best-fit relation is adopted as the albedo input for step 2.

\subsubsection{Step 2: Refined Fit of Phase Function Parameters}

We repeat the fitting procedure from step 1, this time fixing the albedo to the best-fit wavelength-dependent line from Figure \ref{fig:comparing_albedos}. We again use \texttt{emcee} to sample the posterior distributions of the six phase function parameters and $C(\lambda)$. For each filter, we use 25 walkers for 4,000 iterations, with a burn-in period of 3,000 steps. Most filters converge within 500 iterations. No additional priors are applied beyond the normalization requirement for the phase function. The median values of the posterior distributions of the phase function parameters are shown in Figure~\ref{fig:trends_in_g_final}.

\subsubsection{Step 3: Wavelength Dependence of Phase Function Parameters}

Finally, we use \texttt{emcee} to fit wavelength-dependent trends to the six phase function parameters from step 2, yielding a continuous phase function across 0.3–1.6 \micron\ that can be evaluated at any wavelength in this range. We assume a linear relation with wavelength for $g_1$, $g_2$, $w_1$, and $w_2$. We assume constant values for $g_3$ and $w_3$ (i.e., wavelength-independent gegenschein components). We again enforce phase function normalization using a Gaussian prior (mean 1, $\sigma=0.001$) and sample the following likelihood:

\begin{equation}\label{eq:likelihood2}
\log \mathcal{L} = -\frac{1}{2} \sum_{j=1}^{25} 
\left[
\frac{(p_{j} - P(\lambda_j))^2}{\sigma_{p,j}^2} + \ln(2\pi \sigma_{p,j}^2)
\right] ,
\end{equation}

\noindent where $p_j$ is the median posterior value from step 2 for filter $j$, $\sigma_{p,j}$ is the 68.27\textsuperscript{th} percentile width of the posterior, and $P(\lambda_j)$ is the linear-fit model evaluated at the pivot wavelength for the filter $\lambda_j$. We use 30 walkers for 3,000 iterations with a burn-in of 1,000. Most parameters converge within 500 iterations.

\begin{figure*}
    \centering
    \includegraphics[width=\linewidth]{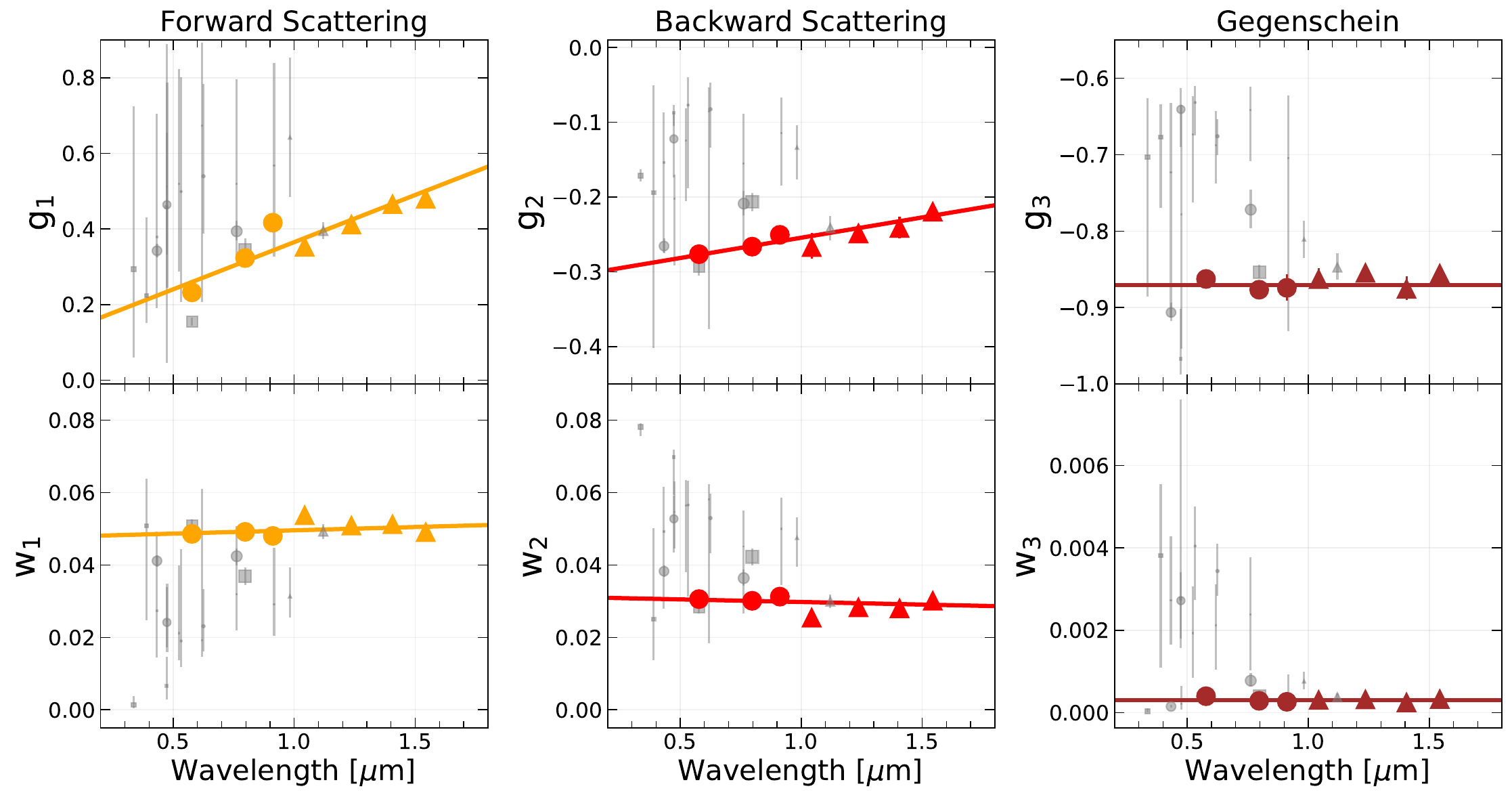}
    \caption{Fitting of phase function parameters to SKYSURF data: forward scattering (left), backward scattering (center), and gegenschein (right). Each point represents the best-fit for that filter, where the albedo is fixed following Figure \ref{fig:comparing_albedos}. The error bars represent the 68.27th-percentile distribution from the posterior. We fit a line to each parameter (solid colored lines), requiring that the phase function at every wavelength integrates to 1.0. We only perform the linear fitting for filters with more than 300 independent pointings, represented by colored symbols. The grey symbols are sized according to the number of unique pointings in that filter (Table \ref{tab:sky_catalog_table}). We assume that the gegenschein components do not change significantly with wavelength, and instead just fit a median. Different symbol shapes represent different HST cameras: squares represent WFC3/UVIS, circles represent ACS/WFC, and triangles represent WFC3/IR.}
    \label{fig:trends_in_g_final}
\end{figure*}

The filters with the largest numbers of independent pointings generally follow a linear relation, though several filters deviate as outliers. In particular, all WFC3 filters exhibit relatively large uncertainties and fall systematically outside the trend. We restrict this step of the analysis to filters with more than 300 independent pointings (see Section \ref{sec:sky_measurement_catalog}). These are shown as the bold, colored symbols in Figure \ref{fig:trends_in_g_final}, representing those with the lowest uncertainties within their wavelength coverage.


\section{Results}\label{sec:results}

The final albedo values are shown as the solid blue line in Figure \ref{fig:comparing_albedos}. The final phase function parameters are shown as solid lines in Figure \ref{fig:trends_in_g_final}. Both of these figures show the best fits from each HST filter individually, where the solid lines represent linear fits as a function of wavelength. These lines are fit so that this model can be applicable to any optical wavelength. For \obrienzodi, the final albedo and scattering phase function relations are:

\begin{equation}\label{eq:final_albedo}
    \text{Albedo}=0.113\times\lambda \text{[\micron]} + 0.082
\end{equation}
\begin{equation}\label{eq:final_g1}
    g_1 = 0.250\times\lambda \text{[\micron]}+0.116
\end{equation}
\begin{equation}\label{eq:final_g2}
    g_2 = 0.054\times\lambda \text{[\micron]}+-0.309
\end{equation}
\begin{equation}\label{eq:final_g3}
    g_3 = -0.870
\end{equation}
\begin{equation}\label{eq:final_w1}
    w_1 = 0.0018\times\lambda \text{[\micron]}+0.0478
\end{equation}
\begin{equation}\label{eq:final_w2}
    w_2 = -0.0014\times\lambda \text{[\micron]}+0.0312
\end{equation}
\begin{equation}\label{eq:final_w3}
    w_3 = 0.0003
\end{equation}

\noindent\ The model is available as \new{both Python and IDL packages} on the \href{https://github.com/rosaliaobrien/skysurf/}{SKYSURF GitHub repository}, with further information on the public code provided in Appendix \ref{app:public_code}.

\obrienzodi\ is currently limited to HST’s wavelength coverage: 0.3–1.6 \micron. In addition, in this analysis, we exclude HST observations with Sun angles $< 80^\circ$ due to concerns about stray light. Consequently, the model presented here should be considered reliable only for Sun angles $> 80^\circ$.

\subsection{Albedo Results}

We find that a linear relationship between albedo and wavelength provides a good fit to our best-fit albedo values across the HST filters. While this simplified model likely does not capture the true relation between albedo and wavelength, we opted for a linear fit to avoid overfitting. We assume the albedo is uniform across the IPD cloud and consistent across all lines of sight.

Our results are shown in comparison with albedos from several previous studies, including the original Kelsall albedos, in Figure \ref{fig:comparing_albedos}.
\cite{San_2024} present updated albedo estimates using a Bayesian approach that incorporates data from DIRBE, Planck, WISE, and Gaia. Their albedo measurements are larger than both ours and the Kelsall values. \new{When fitting zodiacal light models, the albedo is degenerate with the dust number density, and because \citet{San_2024} adopt a lower dust number density (7.46$\times10^{-8}$ AU$^{-1}$) than assumed in this work (11.3$\times10^{-8}$ AU$^{-1}$), this difference may partially explain the discrepancies seen in Figure \ref{fig:comparing_albedos}.}

We also compare to \cite{Ishiguro_2013}, who used the WIZARD instrument to observe gegenschein in the optical, deriving a geometric albedo of $0.06 \pm 0.01$. \cite{Hanner_1974} estimated albedo by reanalyzing micrometeoroid data from Helios A and Pioneer 10, while \cite{Renard_1995} used the node of lesser uncertainty method to derive albedos in the range of $\sim$0.08–0.15 at 1 AU. \cite{Lumme_1985} used polarimetric observations to estimate a an albedo of 0.04. In general, our albedo measurements agree well with those from  \cite{Ishiguro_2013} and \cite{Renard_1995}, yet fall between the high \cite{Hanner_1974} value and low \cite{Lumme_1985} value.

\new{To compare the wavelength dependence of our albedo measurement, we compare it with published zodiacal light reflectance spectra from \citet{Matsumoto_1996} and \citet{Tsumura_2010}. \citet{Tsumura_2010} used the low-resolution spectrometer aboard the Cosmic Infrared Background ExpeRiment (CIBER) to observe zodiacal light from 0.75–1.6 \micron, and extended their analysis to 2.2 \micron\ using Infrared Telescope in Space (IRTS) data from \citet{Matsumoto_1996}. In both studies, the reflectance spectra were obtained by dividing the measured zodiacal light spectra by an assumed solar spectrum and normalizing at a reference wavelength. Because these measurements represent relative reflectance rather than absolute albedo, the comparison is intended to assess only the spectral shape. In Figure \ref{fig:comparing_albedos}, we therefore plot the reflectance spectra from \citet{Tsumura_2010} as a thick grey line, scaling the amplitude to match the Kelsall albedo at 2.2 \micron. The plotted data are taken from Figure 10 of \citet{Tsumura_2010}. The spectral trends from \obrienzodi\ and CIBER/ IRTS agree very well, although our albedo spectrum has noticeably poorer spectral resolution.}

Finally, we include modeled albedos for interstellar dust from \cite{Zubko_2004}, which were derived by fitting UV–IR extinction, diffuse IR emission, and elemental abundances. Although interstellar dust differs in composition and grain size from IPD, and exhibits features such as PAH emission not yet seen in zodiacal light, we include this to compare interstellar dust and IPD. For example, the IPD particles in our Solar System tend to be $>$10 \micron\ in size \citep{Reach_2003}, compared to interstellar dust that is typically $<$1 \micron\ in size \citep{Weingartner_2001}. The models from \cite{Zubko_2004} find that interstellar dust is much more reflective than the dust in our model, but still find a decreasing albedo with decreasing wavelength at the bluest wavelengths.

\subsection{Phase Function Results}

The six phase function parameters are the weight parameters ($w_1, w_2, w_3$) and asymmetry parameters ($g_1, g_2, g_3$). To summarize, $g_1$ and $w_1$ describe forward scattering, $g_2$ and $w_2$ capture backward scattering, and $g_3$ and $w_3$ represent the gegenschein component.

We find that a linear relationship between each phase function parameter and wavelength provides a good fit to the best-fit values derived from our HST observations. The final linear relations, shown as solid lines in Figure \ref{fig:trends_in_g_final} and parametrized in Equations \ref{eq:final_g1}--\ref{eq:final_w3}, are used in \obrienzodi. As described in Section \ref{sec:fitting}, we normalize the final phase function such that its integral over all scattering angles equals 1, which is accurate to within 0.003 (or 0.3\%) in the final fit. 

\begin{figure}[ht]
    \centering
    \includegraphics[width=\linewidth]{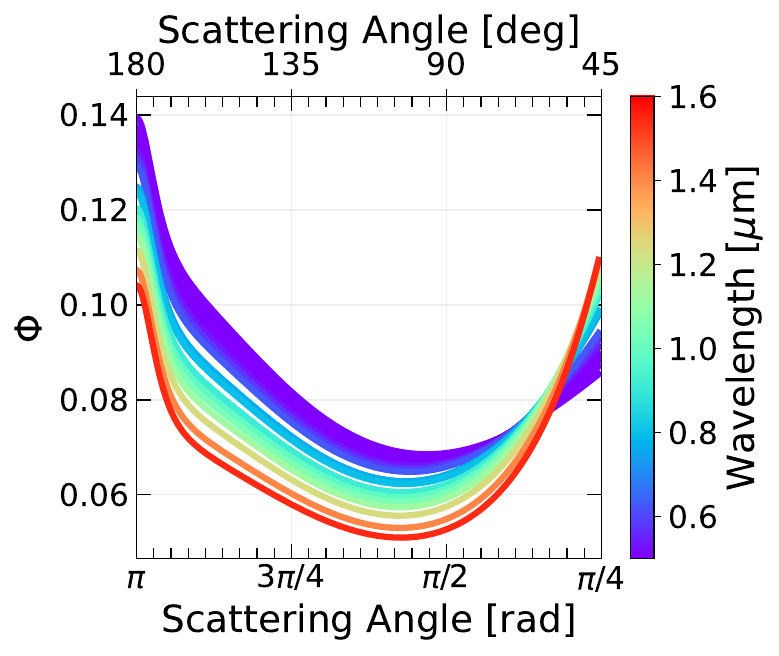}
    \caption{Scattering phase function ($\Phi$) for \obrienzodi, shown for various filter wavelengths (colored lines). The x-axis is limited to 0.8 radians ($\sim$45\degree), just below the minimum Sun Angle observable by HST.}
    \label{fig:phasefunc_colored_wave}
\end{figure}

The final phase function parameters show a clear wavelength dependence, which results in noticeable changes to the shape of the scattering phase function. This is illustrated in Figure \ref{fig:phasefunc_colored_wave}. Specifically, forward scattering becomes more prominent at longer wavelengths, as seen in the increasing trend of $g_1$, which changes more significantly than $g_2$. 

\subsection{Uncertainties in \obrienzodi}\label{sec:uncertainties}

In this section we estimate both the statistical and systematic uncertainties in \obrienzodi. To estimate the statistical uncertainty in \obrienzodi, we use the posterior distributions from step 1 of our MCMC fitting pipeline. This step jointly fits both the albedo and the six phase function parameters. We use posteriors from this step because it includes the largest number of free parameters. In later steps, either the albedo or phase function is held fixed, which could lead to underestimated uncertainties if those steps were used instead.
For each HST filter, we estimate the statistical uncertainty ($\sigma_{\mathrm{stat}}$) in our model using results from the final 1,000 MCMC steps of each of the 25 walkers, totaling 25,000 samples. At each of these 25,000 iterations, we generate model predictions for 100 simulated sky pointings, each with a randomly selected sky position and day of year. This produces 25,000 zodiacal light intensity values for each of the 100 sky pointings. For each pointing, we compute the inner 68.27th percentile of these values, which we define as the 1$\sigma_{\mathrm{stat}}$ uncertainty for that specific pointing. Finally, we take the median uncertainty across all 100 sky pointings as the representative 1$\sigma_{\mathrm{stat}}$ uncertainty for the given HST filter.

We also consider any systematic uncertainty ($\sigma_{\mathrm{sys}}$) in our model. In \citetalias{OBrien_2023}, we present uncertainties in measurements of the sky-SB as random uncertainties regarding the ability of the standard HST calibration pipeline to determine bias frames, dark frames, flat fields, and the photometric zeropoint. \citetalias{OBrien_2023} assumes that the various HST instrument science reports properly correct for subtle systematic offsets in these calibration factors. Nonetheless, Appendix D.6 of \citetalias{OBrien_2023} also discovers that there are low-level flat field residuals present in HST images, and residuals such as these can contribute to systematic offsets in \obrienzodi. In this paper, we discover systematic differences between the WFC3 and ACS detectors for similar bandpasses, reinforcing that there are subtle remaining systematic offsets present in SKYSURF sky-SB measurements, although these are small ($\lesssim0.005$ \MJy). The two main sources of uncertainty in sky-SB measurements that may contribute to the systematic differences are the photometric zeropoints and the flat-fields. Therefore, for this work, we will include the sky-SB uncertainties from \citetalias{OBrien_2023} to be the systematic uncertainties in \obrienzodi. The total \obrienzodi\ systematic uncertainty is 3\% of the solar irradiance spectrum added in quadrature with the sky-SB uncertainty from \citetalias{OBrien_2023}. 

Table \ref{tab:resid_vs_wave_table} shows the resulting uncertainties for each HST filter. The median statistical uncertainty across all HST filters is 0.0009 \MJy. For a zodiacal light level of $\sim$0.1 \MJy (typical in the ecliptic poles around 0.6 \micron, see Figure \ref{fig:comparing_to_other_models}), this results in a random uncertainty of $<1$\%. Filters with the most sky coverage tend to have the smallest random uncertainties. Systematic uncertainties dominate \obrienzodi\ uncertainties, and range from $\sim$0.002 to 0.007 \MJy.  The systematic uncertainties increase with wavelength, largely due to the flat field uncertainty being a multiplicative uncertainty on the total sky-SB ($\sim$1\% of the sky-SB for ACS/WFC, for example).

When considering all HST pointings used in this work, the median \obrienzodi\ uncertainty ($\sigma_{\mathrm{stat}} + \sigma_{\mathrm{sys}}$$ =\sigma_{\mathrm{\obrienzodi}}$) is $\sim$4.5\% of the zodiacal light intensity. It roughly follows
\begin{equation}
    \sigma_{\mathrm{\obrienzodi}} = 0.001\times \lambda + 0.005 \text{ \MJy},
\end{equation}
\noindent where $\lambda$ is in units of \micron. We recommend users of the model assume this uncertainty at $0.3<\lambda<1.6$.

\begin{table*}[ht]
    \caption{Statistical \& systematic uncertainties, and diffuse light residual from \obrienzodi. The first three columns list the HST camera, filter, and pivot wavelength. The fourth and fifth columns list the statistical and systematic uncertainties for \obrienzodi, respectively. In the sixth column, we list the residual diffuse light brightness (HST Sky -- DGL -- \obrienzodi), and the final column lists its standard deviation. Pointings within $20\deg$ of the ecliptic plane are ignored for diffuse light measurements.}
    \centering
    \begin{tabular}{c|c|c|c|c|c|c}
    Camera & Filter &
    \shortstack{Wavelength\\$[$\micron$]$ } & \shortstack{$\sigma_{\mathrm{stat}}$\\$[$\MJy$]$} & \shortstack{$\sigma_{\mathrm{sys}}$\\$[$\MJy$]$} &
    \shortstack{Diffuse Light\\$[$\MJy$]$} & \shortstack{$\sigma_{\mathrm{DL}}$\\$[$\MJy$]$}\\
    \hline
    \hline
WFC3/UVIS & F336W & 0.34 & 0.0002 & 0.0046 & 0.0001 & 0.0092 \\
WFC3/UVIS & F390W & 0.39 & 0.0009 & 0.0030 & 0.0031 & 0.0035 \\
WFC3/UVIS & F438W & 0.43 & 0.0014 & 0.0056 & 0.0022 & 0.0056 \\
ACS/WFC & F435W & 0.43 & 0.0002 & 0.0022 & 0.0024 & 0.0054 \\
WFC3/UVIS & F475W & 0.47 & 0.0009 & 0.0031 & 0.0084 & 0.0031 \\
ACS/WFC & F475W & 0.47 & 0.0005 & 0.0032 & 0.0070 & 0.0042 \\
WFC3/UVIS & F475X & 0.47 & 0.0012 & 0.0038 & 0.0099 & 0.0039 \\
WFC3/UVIS & F555W & 0.52 & 0.0018 & 0.0046 & 0.0151 & 0.0044 \\
ACS/WFC & F555W & 0.53 & 0.0008 & 0.0038 & 0.0077 & 0.0028 \\
ACS/WFC & F606W & 0.58 & 0.0003 & 0.0037 & 0.0124 & 0.0039 \\
WFC3/UVIS & F606W & 0.58 & 0.0002 & 0.0036 & 0.0159 & 0.0030 \\
WFC3/UVIS & F625W & 0.62 & 0.0017 & 0.0051 & 0.0213 & 0.0297 \\
ACS/WFC & F625W & 0.63 & 0.0009 & 0.0052 & 0.0145 & 0.0069 \\
WFC3/UVIS & F775W & 0.76 & 0.0030 & 0.0079 & 0.0193 & 0.0038 \\
ACS/WFC & F775W & 0.76 & 0.0006 & 0.0050 & 0.0135 & 0.0059 \\
ACS/WFC & F814W & 0.80 & 0.0002 & 0.0050 & 0.0137 & 0.0070 \\
WFC3/UVIS & F814W & 0.80 & 0.0005 & 0.0057 & 0.0190 & 0.0071 \\
ACS/WFC & F850LP & 0.91 & 0.0002 & 0.0069 & 0.0168 & 0.0069 \\
WFC3/UVIS & F850LP & 0.92 & 0.0096 & 0.0166 & 0.0203 & 0.0130 \\
WFC3/IR & F098M & 0.98 & 0.0028 & 0.0080 & 0.0015 & 0.0071 \\
WFC3/IR & F105W & 1.04 & 0.0008 & 0.0062 & 0.0095 & 0.0170 \\
WFC3/IR & F110W & 1.12 & 0.0012 & 0.0060 & 0.0118 & 0.0056 \\
WFC3/IR & F125W & 1.24 & 0.0007 & 0.0064 & 0.0122 & 0.0046 \\
WFC3/IR & F140W & 1.41 & 0.0017 & 0.0065 & 0.0161 & 0.0069 \\
WFC3/IR & F160W & 1.54 & 0.0008 & 0.0074 & 0.0173 & 0.0094 \\
\hline
    \end{tabular}
    \label{tab:resid_vs_wave_table}
\end{table*}

\subsection{Diffuse Light}

Diffuse light is the residual light after subtracting foregrounds: 

\begin{equation}\label{eq:diffuselight_diff}
    \text{Diffuse Light} = \text{HST Sky} - \text{DGL} - \obrienzodi
\end{equation}

\noindent This quantity is not necessarily the same as $C(\lambda)$ in Equation \ref{eq:observed_zodi}, since $C(\lambda)$ is allowed to vary independently for each filter during steps 1 and 2 of the fitting procedure. In contrast, the diffuse light defined above represents the final residual signal resulting from the linear trends (step 3 of the fitting procedure) shown in Figures \ref{fig:comparing_albedos} and \ref{fig:trends_in_g_final}.

\begin{figure*}[t]
    \centering
    \includegraphics[width=\linewidth]{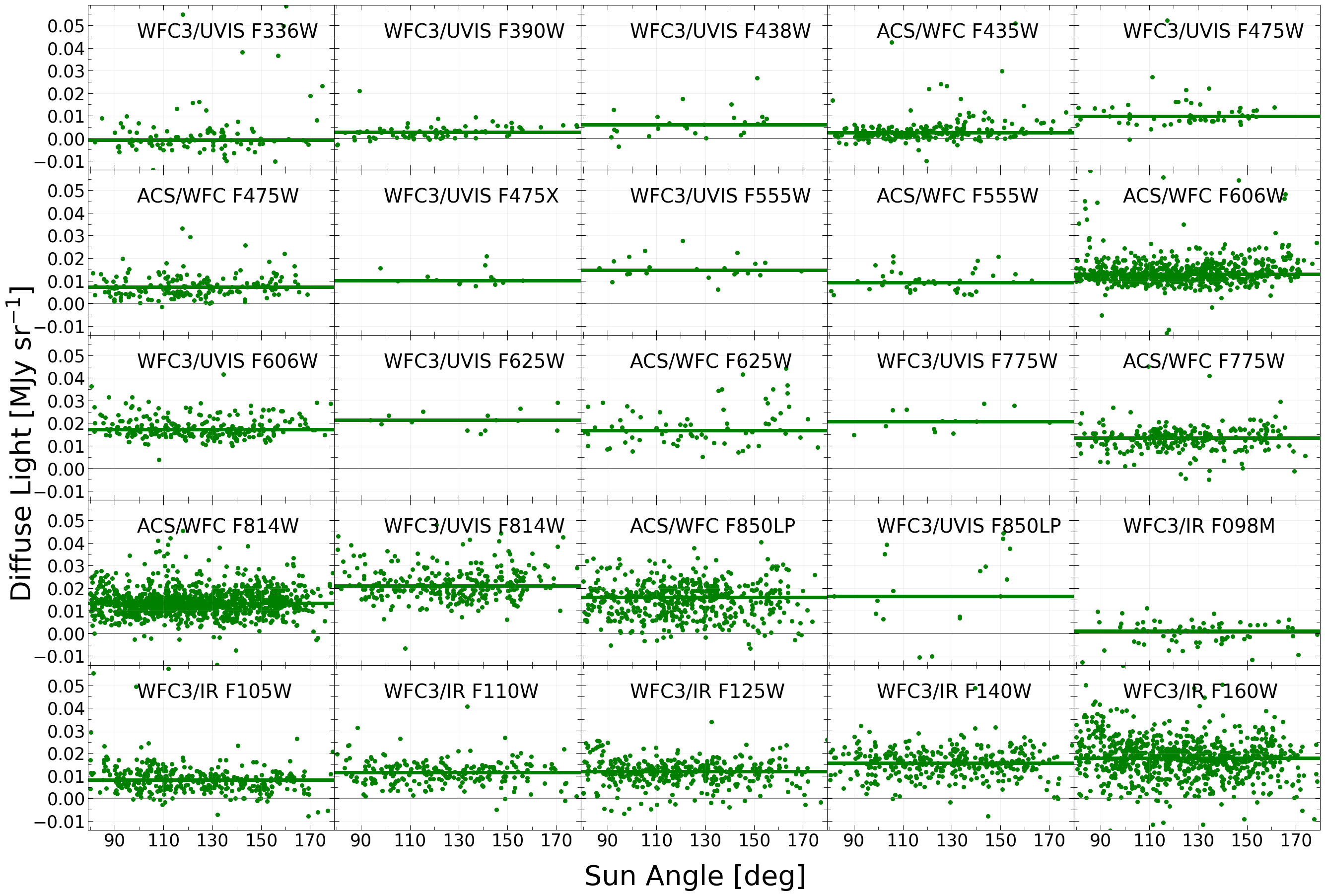}
    \caption{Diffuse light brightness (HST Sky $-$ DGL $-$ \obrienzodi)
    versus Sun angle. Each green point 
    represents an individual HST measurement. The solid green lines 
    show median offsets. The flat residuals indicate the assumed scattering physics is appropriate. HST camera and filter names are shown in black, and the figure panels are sorted from shortest (top-left) to longest (bottom-right) wavelength.}\label{fig:resid_vs_sunang}
\end{figure*}

\begin{figure*}[t]
    \centering
    \includegraphics[width=\linewidth]{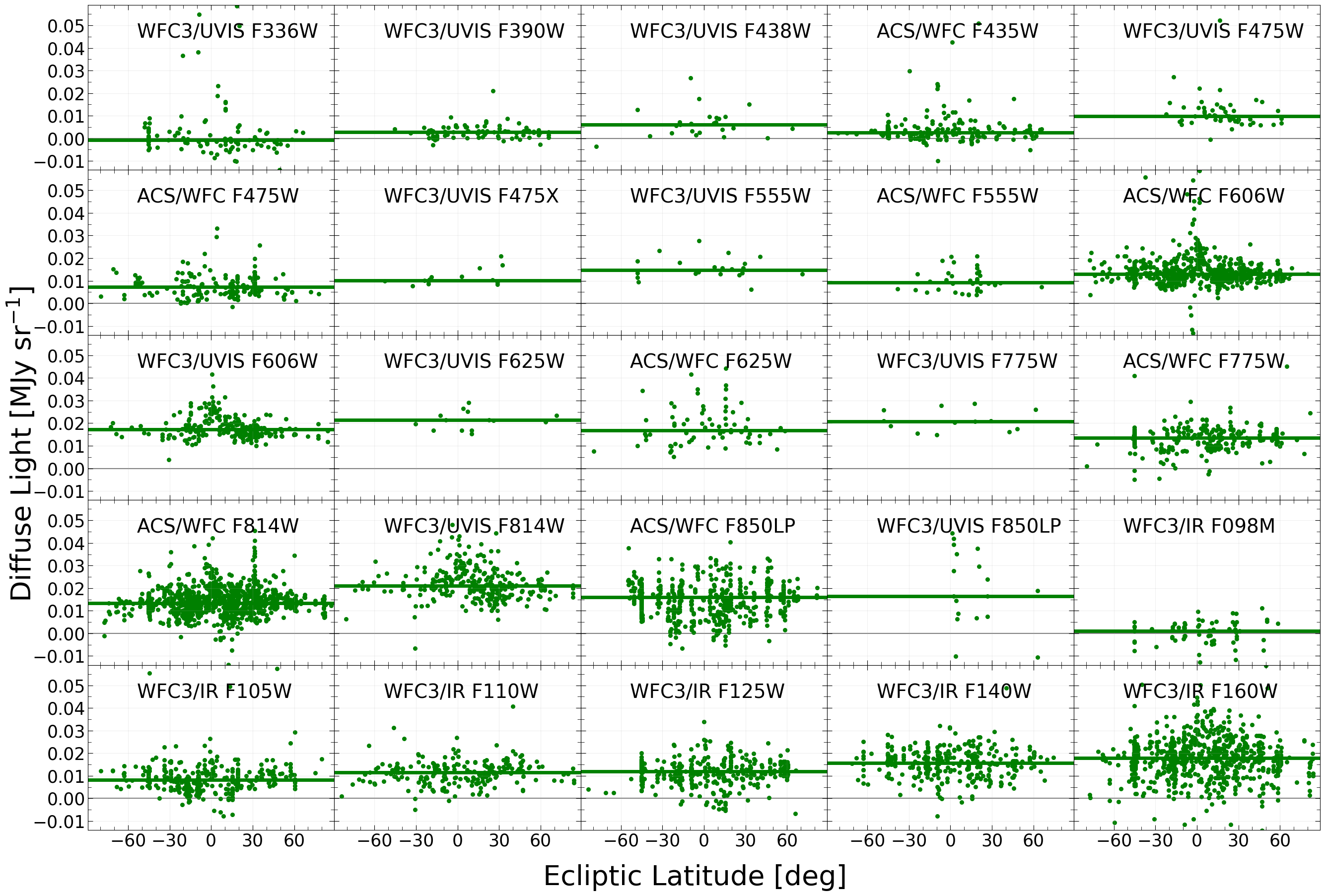}
    \caption{As Figure \ref{fig:resid_vs_sunang}, but plotted versus ecliptic latitude, probing various regions of the IPD cloud. Here, flat residuals indicate that the assumed spatial distribution of IPD is appropriate.}
    \label{fig:resid_vs_ecl_lat}
\end{figure*}

\begin{figure*}[t]
    \centering
    \includegraphics[width=0.9\linewidth]{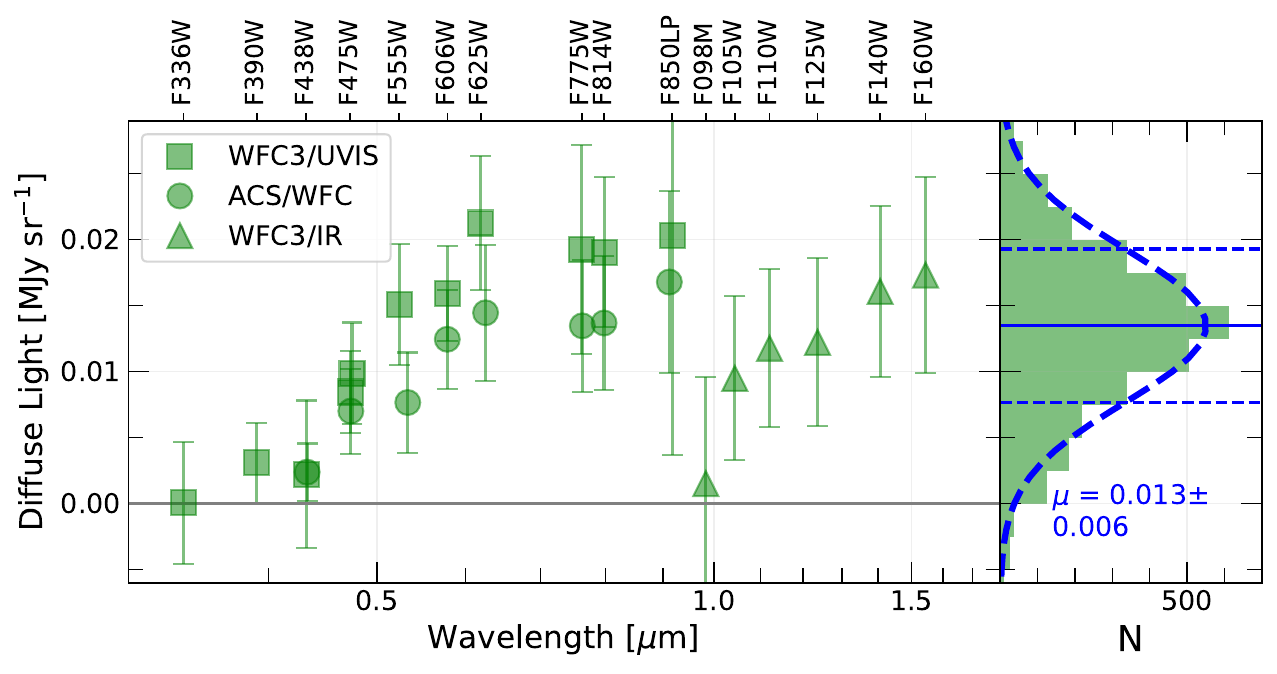}
    \caption{Comparison of diffuse light levels (HST Sky $-$ DGL $-$ \obrienzodi) for each HST filter, as a function of wavelength. Each point represents an average diffuse light brightness for that filter. Different symbol shapes represent different HST cameras: squares represent WFC3/UVIS, circles represent ACS/WFC, and triangles represent WFC3/IR. The errorbars represent the median error in \obrienzodi\ for that filter (Table \ref{tab:resid_vs_wave_table}). The histogram on the right shows the distribution of diffuse light measurements for individual HST pointings, specifically for filters with at least 300 independent pointings. In blue text, we show the mean and standard deviation of this distribution, when a Gaussian profile is fit to it. Pointings within $20\degree$ of the ecliptic plane are ignored. The top axis lists the HST filter corresponding to each diffuse light measurement.
    }
    \label{fig:resid_vs_wave}
\end{figure*}

\citetalias{Carleton_2022}, \citetalias{OBrien_2023}, and \citetalias{McIntyre_2025} report an isotropic diffuse light component of $\sim$0.01 \MJy\ in HST measurements, so we expect some residual diffuse emission even after subtracting \obrienzodi\ from the sky-SB data. A small fraction of this residual arises from unresolved galaxies fainter than our AB$\sim$27 mag limit, with the rest remaining unexplained. In Figures \ref{fig:resid_vs_sunang}--\ref{fig:resid_vs_wave}, we analyze the diffuse light estimate of Equation~\ref{eq:diffuselight_diff} using our \obrienzodi\ model prediction, our DGL estimator, and our HST sky-SB measurements. We perform our analysis of diffuse light while ignoring all data within $20\deg$ of the ecliptic plane, due to brighter DGL in this region.

Regardless of whether diffuse light averages to zero, its trends with Sun angle and ecliptic latitude can test the fidelity of \obrienzodi\ (\eg\ excess brightness at certain Sun angles may indicate an inaccurate scattering phase function).
Figures \ref{fig:resid_vs_sunang} and \ref{fig:resid_vs_ecl_lat} show the residual diffuse light signal for our \obrienzodi\ model compared to our HST sky-SB measurements, as a function of Sun angle and ecliptic latitude, respectively. The \obrienzodi\ model demonstrates flat residuals for all wavelengths. The flat residual as a function of Sun angle indicates that the scattering physics in the updated model agrees well with HST observations, as a range of Sun angles also probes a range of scattering angles. The flat residuals as a function Sun angle also indicate that there is no stray light entering HST's CCD, as stray light should become more significant at smaller Sun angles.

The flat residual as a function of ecliptic latitude indicates that the three-dimensional structure of the cloud is accurate. However, there is more scatter at lower ecliptic latitudes, indicating some component of the model may be improved at low ecliptic latitudes. For example, the average standard deviation of diffuse light measurements within 30\degree\ of the ecliptic plane is $\sim$25\% larger than outside of the ecltipic plane. Most obviously, the model underpredicts the brightness for the ACS/WFC and WFC3/UVIS F606W filters at low ecliptic latitudes. \cite{San_2024} find a faint band along the ecliptic plane when re-examining DIRBE data, and state that high-resolution measurements of zodiacal light along the ecliptic plane may help distinguish an additional component in this part of the ecliptic. For this reason, for the diffuse light analysis in this paper, we ignore HST pointings within 20\degree\ of the ecliptic plane.


Figure \ref{fig:resid_vs_wave} shows our broadband diffuse light spectral energy distribution, and Table \ref{tab:resid_vs_wave_table} lists the diffuse light measurement and standard deviation ($\sigma_{\mathrm{DL}}$) for each HST filter. Our diffuse light measurements rise smoothly from 0 to $\sim$0.013 \MJy\ between 0.3–0.6 \micron, then flatten beyond 0.6 \micron.
Diffuse light levels at $\sim$0.013 \MJy\ is $\sim$$2\times$ more than the predicted levels of IGL, indicating that either the foreground components are not modeled accurately (i.e., possibly missing components to DGL or \obrienzodi), or there is a significant missing (diffuse) source population in EBL models. If the diffuse light signal came from a missing zodiacal light component, this could explain the systematic offset in our model. Further discussion on whether our measured diffuse light may be from IPD is included in Section \ref{sec:spherical_comp}. 



\subsection{Comparison with Other Models}

Figure~\ref{fig:comparing_to_other_models} compares our total sky-SB model (\obrienzodi\ $+$ DGL) with \citetalias{OBrien_2023}, \new{other direct measurements \citep{Giavalisco_2002, Kawara_2017, Matsuura_2017},} and several existing zodiacal light models \citep{Kelsall_1998, Reach_1997, Wright_1998, Aldering_2001, San_2024}. This comparison is restricted to observations taken within 45\degree\ of the ecliptic poles, where the sky is darkest.

Both the Kelsall and Wright models have been implemented into the IPAC IRSA Background Model\footnote{\url{https://irsa.ipac.caltech.edu/applications/BackgroundModel/}}, for wavelengths as short as 0.5 \micron. Since this wavelength is considerably shorter than the shortest nominal wavelength of the DIRBE instrument, the IPAC IRSA Background Model must make assumptions about how zodiacal light predictions extend to $\lambda < 1.25$ \micron. For this discussion, we will therefore refer to these models as the IPAC-Kelsall model and the IPAC-Wright model. We also compare to the The Cosmoglobe ZodiPy model \citep{San_2024} and the JWST Background Tool\footnote{\url{https://jwst-docs.stsci.edu/jwst-other-tools/jwst-backgrounds-tool}}. ZodiPy is a Python package designed to remove foreground contamination for the Cosmoglobe project. The JWST Background Tool was developed to predict zodiacal light at L2, and currently is a modified version of the IPAC IRSA Background model. It is known to overpredict at JWST's bluest wavelengths \citep{Rigby_2023}.  The \cite{Aldering_2001} model is empirically derived to match observations in the ecliptic pole and is based on a reddened solar spectrum and a simplified dust cloud.

\begin{figure}
    \centering
    \includegraphics[width=\linewidth]{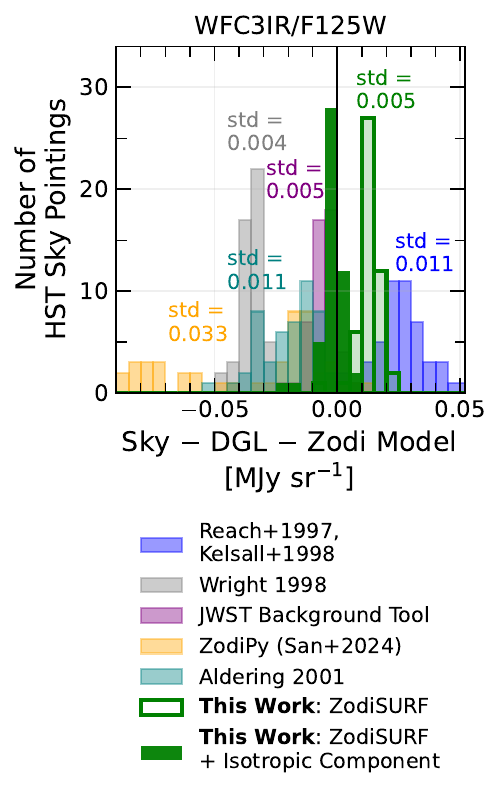}
    \caption{Comparison of the distribution of diffuse light measurements with \obrienzodi\ and other zodiacal light models. The diffuse light is HST Sky -- DGL -- Zodi Model for each HST sky-SB measurement at 1.25 \micron. The DGL model is the same for all zodiacal light models. The models are the same as described in Figure \ref{fig:comparing_to_other_models}. In colored text within the figure, we show the standard deviation of each distribution.}
    \label{fig:comparing_distribution_models_125m_pretty}
\end{figure}

\begin{figure}[t]
    \centering
    \includegraphics[width=\linewidth]{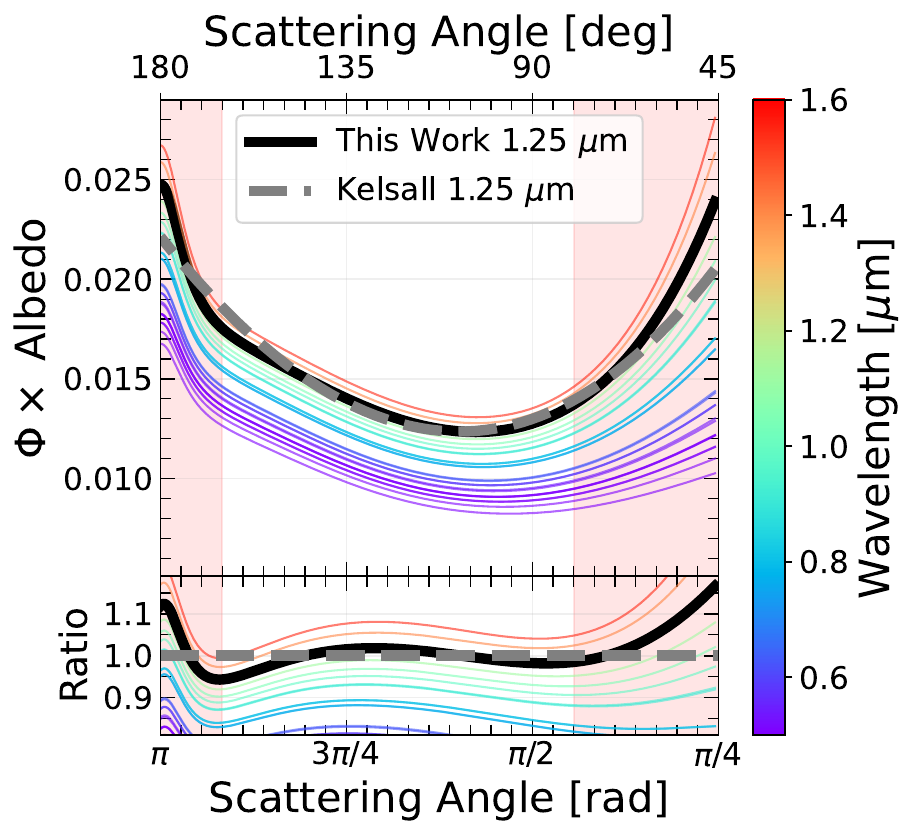}
    \caption{Scattering phase function ($\Phi$) times albedo for the \obrienzodi\ model, shown for various filter wavelengths (colored lines), compared to the original Kelsall model (grey dashed line). We highlight $\Phi\times$Albedo for $\sim$1.25 \micron\ (WFC3/IR F125W filter) from this work as a dark black line for comparison with the Kelsall model. HST did not observe at Sun Angles $<$ 80\degree, and therefore \obrienzodi\ cannot constrain at smaller angles (red shaded region on the right side). COBE did not observe at Sun Angles $>$ 124\degree, and therefore the Kelsall model was largely insensitive to the gegenschein region (red shaded region on the left side, which ranges from 165\degree to 180\degree). The top panel shows the absolute values of $\Phi\times$Albedo, while the bottom panel shows the ratio between \obrienzodi\ and Kelsall for each wavelength. The x-axis is limited to 0.8 radians ($\sim$45\degree), just below the minimum Sun Angle observable by HST.}
    \label{fig:phasefuncTalbedo_colored_wave_wkelsall}
\end{figure}

What makes \obrienzodi\ a transformative model is its ability to reliably predict zodiacal light emission at any wavelength between 0.3--1.6 \micron. As Figure~\ref{fig:comparing_to_other_models} shows, \obrienzodi\ is the only current model with this capability. For example, at 0.6 \micron, \obrienzodi\ brings the total sky brightness prediction 75\% closer to the measured HST value when compared to the JWST Background Tool.
Except the Aldering model, all models tend to overpredict at  $\lesssim1$ \micron. The IPAC-Wright model consistently overpredicts the sky brightness across all wavelengths. The ZodiPy model and JWST Background Tool show similar performance to the IPAC-Kelsall model, likely because both adopt similar assumptions about scattering at optical wavelengths. The Aldering model tends to overestimate the sky brightness slightly in the near-IR ($>$1 \micron), but it is relatively accurate at optical wavelengths. However, this model does not accurately capture spatial variations in dust geometry (\eg\ dust bands or circumsolar ring) or more complex scattering physics (\eg\ gegenschein).

\new{We also show direct measurements of zodiacal light from \cite{Kawara_2017} and \citet{Matsuura_2017}.} \cite{Kawara_2017} decompose the sky brightness into zodiacal light, DGL, and residual emission using the Faint Object Spectrograph on board HST. These measurements are estimated for an ecliptic latitude of 85$\degree$ using Table 2 and Equation (8) from \cite{Kawara_2017}. 
\new{
\citet{Matsuura_2017} use the 0.8--1.7 \micron\ spectra from CIBER to measure EBL by subtracting measurements of zodiacal light, stars, terrestial emission, and DGL. The plotted spectrum is a weighted mean from the three CIBER flights, and is scaled to match our \obrienzodi\ prediction at 1.25 \micron. The zodiacal light measurements from both papers agree well with our \obrienzodi\ $+$ DGL predictions.
}

To highlight the accuracy of each model (\eg\ for a range of Sun Angles, positions, and dates), we show the distribution of diffuse light measurements for each of the aforementioned zodiacal light models in Figure \ref{fig:comparing_distribution_models_125m_pretty}. \obrienzodi\ is one of the most precise models (a standard deviation in diffuse light of 0.005 \MJy\ at 1.25 \micron), where only the JWST Background Tool and Wright models have comparable precision. Still, the Wright model significantly overestimates zodiacal light for our HST sky pointings. The JWST Background model predicts the zodiacal light foreground well at 1.25 \micron, but tends to overestimate at $\lambda<1.0$ \micron. 


Overall, our model of the full sky-SB (\obrienzodi\ $+$ DGL) matches the sky-SB measurements well at $0.2<\lambda<1.6$ \micron\ in the ecliptic pole regions. Aside from \obrienzodi, the only model that accurately matches the spectral shape of direct sky-SB measurements is the Aldering model, but due to its simplified assumptions, will result in noticeably lower accuracy across a range of positions and dates, as shown in Figure \ref{fig:comparing_distribution_models_125m_pretty}.

We also compare \obrienzodi\ to the original Kelsall model at 1.25 \micron. This is not the IPAC-Kelsall implementation, but rather the original code as is. The Kelsall model is considered reliable at 1.25 \micron, so our model, with our independently constrained phase function and albedo, should match the original model closely. Figure \ref{fig:phasefuncTalbedo_colored_wave_wkelsall} plots the product of the phase function and the albedo (essentially the effective scattering efficiency) alongside that from Kelsall. Because the albedo and the phase function are strongly correlated, changes in one can be offset by compensating changes in the other. Comparing their product therefore offers a fair, one-to-one assessment of how our updated zodiacal light model differs from Kelsall across wavelength.

The phase function results in this work agree well with Kelsall at intermediate scattering angles (110\degree\ $\pm$ 40\degree). The most noticeable differences are the inclusion of gegenschein for backward scattering, and more forward scattering than the Kelsall model. COBE was limited to Sun Angles between 64\degree and 124\degree. This, and the fact that the Kelsall model cuts off at 5.2 AU, means that the Kelsall model was very insensitive to the gegenschein component. For example, with a limit of $\epsilon<124\degree$, COBE did not see any scattering at angles from $\gtrsim165\degree$ from dust within 3.3 AU of the Sun. In Figure \ref{fig:phasefuncTalbedo_colored_wave_wkelsall}, we highlight in red the region where HST and COBE were limited due to Sun angle constraints ($\epsilon<80\degree$ for HST and $\epsilon>165\degree$ for COBE), and therefore the comparison is unreliable at these scattering angles. A continued discussion on differences between the models, including two-dimensional maps of the ratio and difference between both models at 1.25 \micron, is shown in Appendix \ref{app:comparing_hst_and_cobe}. Overall, our model agrees sufficiently with the Kelsall model at 1.25 \micron.

\section{Discussion}\label{sec:discussion}

Figure \ref{fig:comparing_to_other_models} showcases how \obrienzodi\ improves zodiacal light modeling between 0.3--1.6 \micron. We now discuss how our albedo and scattering phase function results inform the composition of the IPD, explore possible sources of diffuse light, and consider the potential existence of an additional, isotropic zodiacal light component.

\begin{figure*}[t]
    \centering
    \includegraphics[width=\linewidth]{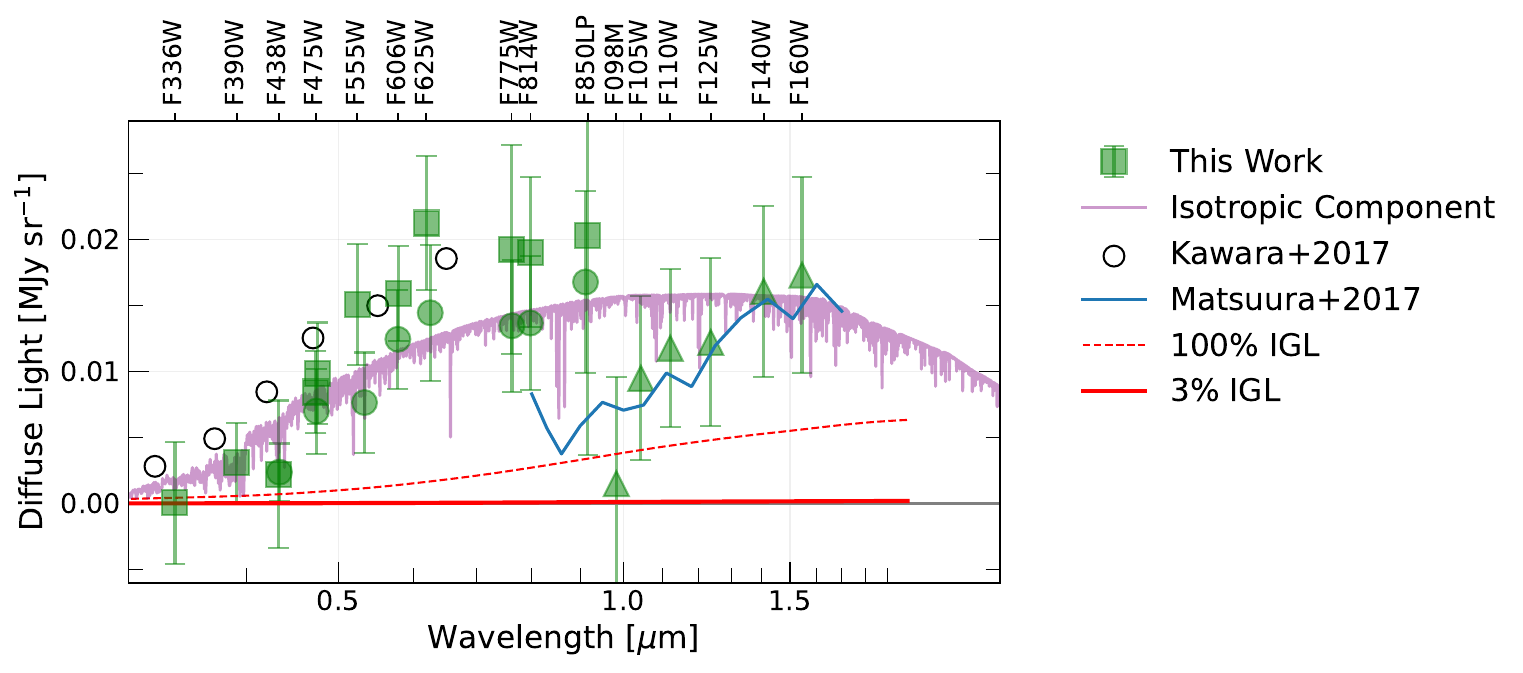}
    \caption{Same as Figure \ref{fig:resid_vs_wave}, but compared with other spectra and measurements of diffuse light.  The purple solid line represents the spectrum of a proposed isotropic component to zodiacal light (see Section \ref{sec:spherical_comp}).
    \cite{Kawara_2017} estimated diffuse light using the Faint Object Spectrograph on board the Hubble Space Telescope, which we plot as open black circles.
    \new{\cite{Matsuura_2017} measure EBL using CIBER spectra, and we plot their estimate for diffuse light as a solid blue line by subtracting an IGL prediction \citep{Keenan_2010} from their nominal EBL measurement.}
    We plot the the total EBL estimate from \cite{DSilva_2023} as a dashed red line. However, our estimates of diffuse light already have $>$$97\%$ of EBL removed. Therefore, we also show the amount of EBL still present in our estimates (3\% EBL) as a solid red line. 
    }
    \label{fig:resid_vs_wave_comparison}
\end{figure*}

\subsection{Implications on the IPD Composition \& Size Distribution}

The albedo and phase function measurements from this work provide insights into the properties of the IPD cloud, although our interpretations are limited since we used broadband imaging with minimal spectral resolution. We fit the albedo independently for each HST filter and find that it increases with wavelength. Our model assumes a simple, linear increase, though this is likely an oversimplification, due to varying dust compositions and sizes. \citet{Tsumura_2010} and \citet{Kawara_2017} derived zodiacal dust “reflectance” spectra without explicitly incorporating a detailed scattering phase function into their fits. The spectra from \cite{Tsumura_2010} \new{(and updated results from \citeauthor{Matsuura_2017} \citeyear{Matsuura_2017})} follow a similar wavelength dependence as our albedo, although their trend is non-linear, highlighting the importance of spectral data for improved zodiacal light modeling. Meanwhile, \citet{Kawara_2017} report a pronounced dip in reflectance at $\sim$0.3 \micron, which they attribute to intrinsic dust properties; for example, \citet{Matsuoka_2015} found a similar dip in the reflectance of meteorite samples from C-type (``chondrite''; clay and silicate) asteroids. We do not observe such a dip in our data, but these results still suggest that albedo is unlikely to vary linearly with wavelength, as assumed in our model.

Still, the measured albedos can offer clues about the composition of the IPD grains.
The dynamical analysis from \citet{Nesvorny_2010} suggests that IPD grains consist primarily of cometary dust, which generally have lower albedos than asteroidal dust.
For context, comets have geometric albedos $p_V<0.05$ \citep[see Figure 1 of ][]{Yang_2015}, while S-type (silicate and nickel-iron) asteroids have much higher geometric albedos ($p_V\sim0.2$) \citep{Tedesco_1989}. The geometric albedo is different than the albedo used in this work, and Appendix \ref{app:geometric_albedo} discusses the difference and converting between the two. In our case, converting our 1.0 \micron\ albedos to geometric albedos yields $p_V \sim 0.07$. Our measurements agree with the idea that zodiacal light is dominated by scattering from cometary dust. 
For comparison, \citet{Ishiguro_2013} used the Widefield Imager of Zodiacal light with ARray Detector (WIZARD) to study gegenschein in the zodiacal light. For this study, they derive a geometric albedo of the smooth component of the Kelsall model to be $p_V = 0.06 \pm 0.01$. Using these results,
\cite{Yang_2015} (their Figure 6) compare the reflectance of zodiacal light to that of different asteroid types, including D-type asteroids (which have low albedos, red colors, and are likely composed of organic compounds) as analogs to cometary nuclei. They find that the zodiacal light reflectance spectra more closely matches that of D-type asteroids, indicating that $>$90\% of IPD particles originate from comets. However, their reflectance spectra is redder than our albedo spectrum (they have a slope of $\sim$0.4 between 1.0 and 1.5 \micron, which is steeper than our albedo slope of $\sim$0.1). In contrast, based on the shape of their reflectance spectra, \new{\citet{Matsumoto_1996} and} \citet{Tsumura_2010} argue that S-type (silicate and nickel–iron) dust dominates zodiacal light emission seen from Earth, as their measured reflectance closely matches that of S-type asteroids.



The scattering phase function measurements in this work provide clues about the dust sizes in the IPD cloud. We find that forward scattering becomes more pronounced at longer wavelengths, while backward scattering becomes less dominant (Figure \ref{fig:trends_in_g_final}). Following Mie scattering theory, this suggests that the dominant scatterers in the IPD cloud are micron-sized, due to the strong dependence we see with wavelength at $\lambda < 2$ \micron. However, in general, it is expected that zodiacal light is scattered off dust grains with sizes $>10$ \micron. \citet{Hanner_1980}, drawing on lunar microcrater data \citep{Giese_1976, Fechtig_1976}, argue that 80\% of zodiacal light originates from particles larger than 10 \micron.
Similarly, using ISO satellite data, \citet{Reach_2003} concluded that interplanetary dust is dominated by large ($>10$ \micron) particles with low albedos ($<0.08$). \citet{Takimoto_2022} analyze a polarization spectrum of zodiacal light from CIBER and find that it is well reproduced by a Mie scattering model for $>1$ \micron-sized graphite-based grains.
Nonetheless, there is still likely substantial small-grain population. Figure 3 of \citet{Silsbee_2025}, based on Ulysses spacecraft observations \citep{Grun_1997, Wehry_1999}, reveals a significant number of grains with sizes $\lesssim$1 \micron. Using the phase functions from this work to estimate the sizes of IPD grains contributing to zodiacal light would benefit from broader wavelength coverage (up to 4 \micron, to capture the full scattering-dominated regime) and polarization measurements.

Grain size can also be understood in the dynamical context of our Solar System. \cite{Arnold_2019} estimate a blowout radius of approximately 0.8 \micron\ for a solar-type star (with some variation depending on dust composition). The blowout radius marks the grain size at which radiation pressure overcomes gravitational force from the host star, and grains smaller than the blowout size are expelled by stellar radiation pressure. For a blowout radius of 0.8 \micron, particles with radii $\lesssim$1.6 \micron\ are expected to escape on hyperbolic orbits \citep[][]{Krivov_2006}. In contrast, large grains ($\gtrsim$10$\times$ the blowout size) will spiral inward and fall into the Sun, on timescales of about 10,000 years \citep[e.g.,][]{Leinert_1990, Klavcka_2008}. Therefore, based off the results of \cite{Arnold_2019}, the typical grain size would need to be $>$1.6 \micron, which loosely agrees with our preferred dust grain size.



Our measurements show that gegenschein remains relatively flat with wavelength. \citet{Ishiguro_2013} attribute the phenomenon to either coherent backscattering or shadowing effects on rough dust grain surfaces (which obscure light at other phase angles), both of which would require that the dust grain diameter is sufficiently larger than the incoming radiation wavelength. Based on the strength of their gegenschein signal, they conclude that zodiacal light is dominated by particles larger than their observed wavelength of $\sim$0.5 \micron. We detect a significant gegenschein signal that appears to vary little with wavelength, which may suggest that the typical dust grain diameter is in fact $>$2 \micron\ in size.


\subsection{Is Diffuse Light Extragalactic?}


We detect a significant diffuse light signal of $\sim$0.013 \MJy\ at $\lambda \gtrsim0.6$ \micron.
Instrument systematics could theoretically contribute to the diffuse light we measure. The 0.45-0.9 \mum\ ACS/WFC data points (green circles in Figure \ref{fig:resid_vs_wave_comparison}) are consistently below the WFC3/UVIS data points (green squares), highlighting that systematic effects are still present. 
We suspect that any residual WFC3/UVIS--ACS/WFC differences could be due to residual zeropoint errors or dark-frame subtraction errors between these two cameras (for a discussion of these effects, see Section 4 of \citetalias[][]{Windhorst_2022} and Appendices B--E in \citetalias[][]{OBrien_2023}). However, we assume any systematic effects are already included in our uncertainty analysis, and Figure \ref{fig:resid_vs_wave_comparison} shows that it is likely $\lesssim$$0.005$ \MJy.

With the assumption that our diffuse light signal is not an instrumental artifact, we explore possible astrophysical sources of the signal. \new{Diffuse light has been measured before \citep[\eg][]{Cooray_2016, Kawara_2017, Mattila_2017, Sano_2020, Driver_2021, Matsuura_2017, Korngut_2022, Symons_2023}, with origins proposed from missing Solar System components to distant galaxies.} Our work provides the first consistent spectral measurements at wavelengths between 0.3--1.6 \micron, offering new insight into its nature. \new{Our measurements match closely with diffuse light measurements from \cite{Kawara_2017}, who covered 0.2--0.7 \micron, and \cite{Matsuura_2017}, who covered 0.8--1.7 \micron, reinforcing our measurements. For the latter, we estimate their measurements of diffuse light to be their nominal EBL with the IGL from \citet{Keenan_2010} subtracted.}

Possibly the most interesting source of diffuse light would be one of extragalactic origin, as this could mean there were significantly more faint or distant galaxies than current models predict. Figure \ref{fig:resid_vs_wave_comparison} shows the spectrum for IGL from \cite{DSilva_2023}, which is estimated by \citet[][]{Tompkins_2025}. Any IGL still left in the images is ${\sim} 0.56$ \nW, or ${<} 0.0002$ \MJy \citepalias{Carleton_2022}, which is shown as the full-drawn red 3\% IGL line in Figure \ref{fig:resid_vs_wave_comparison}. As explained in \citet[][]{Windhorst_2022} and \citet[][]{Carleton_2022}, the way the object-free sky-SB was measured from the HST images already rejects 97\% of the IGL. As shown in Figure \ref{fig:resid_vs_wave}, our diffuse light measurements cannot be due to IGL. If it were due to IGL, the amount of integrated discrete galaxy light in the universe would have to be underestimated by 200\%. This is very unlikely, as independent methods to estimate IGL in deep HST and JWST imaging find consistent results \citep{Tompkins_2025, Carter_2025, Windhorst_2023}.

\new{Although beyond the scope of this work, studying fluctuations in the EBL can provide additional insight into its complex contributions. For example, large-scale EBL fluctuations exceed predictions based on the IGL \citep[e.g.,][]{Kashlinsky_2012, Cooray_2012}. These excess fluctuations may shed light on the contribution of intrahalo light \citep{Zemcov_2014, Feder_2025} or indicate more exotic sources, such as decaying dark matter or direct-collapse black holes \citep[see discussion in][]{Feder_2025}.}

Lastly, a simple multiplier to \obrienzodi\ (e.g., uniformly increasing its brightness by 5\%) cannot explain the offset, which is nearly constant with ecliptic latitude (Figure \ref{fig:resid_vs_ecl_lat}). The extra signal is isotropic, whereas scaling the model would produce latitude-dependent trends instead of the observed flat relation.

\subsection{A Missing Dim Isotropic Component to Zodiacal Light}\label{sec:spherical_comp}

The existence of a component of the IPD cloud that appears isotropic has been debated for decades. In this work, we define an ``isotropic component'' as any dust distribution that would produce zodiacal light that appears isotropic in all directions when observed from Earth. The simplest physical scenario would be a spherical shell with peak density $>>$1 AU, where changes in Sun angle or ecliptic latitude do not result in noticeable changes in brightness. This shell would need to be faint enough, distant enough, or thin enough such that changes in the line-of-sight column density of the shell result in changes in brightness that are not discernible. The COBE mission (and the Kelsall and Wright models) could not test for any component that appears isotropic, since COBE maps were optimized by mapping temporal changes in sky brightness rather than absolute brightness.

\new{\citet{Matsuura_1995} were the first to suggest such a component based on rocket-borne observations of zodiacal light. They observed that the zodiacal light becomes redder at high ecliptic latitudes and attributed this trend to a possible spherical component of the IPD, similar to the component introduced by \citet{Divine_1993}. Using data from the Pioneer 10 and 11, Helios 1, Galileo, and Ulysses spacecraft, \citet{Divine_1993} developed a model of the interplanetary environment that includes five distinct dust populations with different particle sizes and mass distributions. In particular, the “halo” component of this model could explain the isotropic component proposed here. This halo is characterized by a peak in particle density at 3–5 AU and a nearly uniform distribution of orbital inclination angles. The uniform inclinations are required to reproduce Ulysses measurements \citep{Grun_1992}, which were sensitive to highly inclined orbits. Beyond $\sim$3.4 AU, the particle density remains relatively flat to match Pioneer 10 observations \citep{Humes_1980}, which measured an approximately constant dust flux out to 18 AU.}

\citet{Dwek_1998} argued that the isotropic brightness detected by DIRBE could not originate from a solar system component. However, their Figure 2 illustrates the parameter space, defined by particle size and heliocentric distance, where such a cloud could plausibly exist. They concluded that a cloud located between 5 and 150 AU could remain stable against interactions with the interstellar medium, solar wind, radiation pressure, and solar gravity.

Other studies have continued to explore this possibility. \citet{Rowan-Robinson_2013} modeled IPD emission using IRAS and COBE data, and argued that about 7.5\% of the IPD originates from interstellar sources, which could contribute to an isotropic component. \citet{Sano_2020} analyzed residual light in COBE/DIRBE as a function of solar elongation and found residual signatures consistent with an isotropic component. \citet{Korngut_2022} measured the absolute brightness of zodiacal light using Fraunhofer line spectroscopy with CIBER and reported that their best-fitting model required the addition of an isotropic component to the Kelsall model. The dynamical model from \citet{Nesvorny_2010} shows that Oort-cloud comets could supply a heliocentric isotropic dust population. \citet{Renard_1995} similarly suggested a cometary origin for such a component.

\new{The diffuse light results from \citet{Kawara_2017} and \citet{Matsuura_2017} (shown in Figure \ref{fig:resid_vs_wave_comparison}) are particularly interesting, as both agree with our diffuse light spectrum. \citet{Kawara_2017} find that their spectrum is similar to a zodiacal light spectrum, and argue it could be due to an isotropic component to the IPD cloud. Similarly, \citet{Matsuura_2017} also finds a significant diffuse light excess, which they argue could be an additional foreground component or an additional EBL component (although an isotropic zodiacal light component alone cannot explain their results due to the red color of their spectrum).
However, if indeed all three spectra are real and local, the dip in brightness at $\sim$1 \micron\ would need to be explained. For example, silicate from IPD presents a dip in its reflectance spectra at this wavelength \citep[e.g.,][]{Tsumura_2010}. Specifically, olivine shows strong spectral features at $\sim$1 \micron\ \citep[e.g.,][]{Zeidler_2011}, and is known to be present in the IPD \citep[e.g.,][]{Christoffersen_1986}.}

\new{Nonetheless, not only does this dip occur at the boundary between HST's CCD and infrared detectors, but in addition, out of all SKYSURF IR filters, F098M has by far the poorest statistics, so that the $\sim$1 \micron\ dip could simply be a $\lesssim$2$\sigma$ statistical fluctuation. Missions designed to measure the diffuse sky with sufficient spectral resolution \citep[e.g., CIBER-2 and SPHEREx,][]{Zemcov_2025, Bock_2025} may help confirm our diffuse light spectrum and further reveal its source.}

Observations of extrasolar debris disks provide an external context for isotropic dust components. Some disks show extended structures generated by radiation pressure acting on grains smaller than the blowout radius. A prominent example is Vega \citep{Su_2005}, where collisions between asteroids produce debris that is subsequently blown outward, forming an extended dust population.

To test whether an additional component of the IPD cloud could produce an isotropic component to the zodiacal light emission, we constructed a toy model by adding a spherical component to our baseline model. We defined a simplified dust density profile,

\begin{equation}
    n = n_0 R^{-\alpha},
\end{equation} \label{eq:dust_density}

\noindent where $n_0$ is the column density at 1 AU (AU$^{-1}$), $R$ is the heliocentric radial distance, and $\alpha$ is the power-law slope. \new{The maximum $R$ allowed by the current model is 5.2 AU (the same as used in the original Kelsall model).} If $\alpha < 0$, the dust density increases with distance from the Sun—opposite to the usual expectation that density is highest near the Sun. For comparison, the smooth cloud component in the original Kelsall model has $\alpha = 1.34$. The central question is whether a value of $\alpha$ exists that would yield an apparently isotropic component.

We fit for $n_0$ and $\alpha$ at 1.25 \micron\ using the \texttt{emcee} MCMC sampler to maximize the log-likelihood function

\begin{multline}
\log \mathcal{L}_{1.25} =\\ -\frac{1}{2} \sum_{i=1}^{357} \bigg[
\frac{(\text{Diffuse Light}_i - \obrienzodi_{sph,i})^2}{\sigma_{Z,i}^2} \\ + \ln(2\pi \sigma_{Z,i}^2)
\bigg],
\end{multline}

\noindent for the 357 WFC3/IR F125W HST pointings. $\obrienzodi_{sph,i}$ is a spherical copy of the smooth cloud component of our \obrienzodi\ model, evaluated for the position for HST pointing $i$ at 1.25 \micron. This formulation tests whether our diffuse light measurements can be explained by an isotropic component. In effect, the likelihood function identifies the spherical cloud parameters that best reproduce the observed diffuse light at 1.25 \micron.

\begin{figure}
    \centering
    \includegraphics[width=\linewidth]{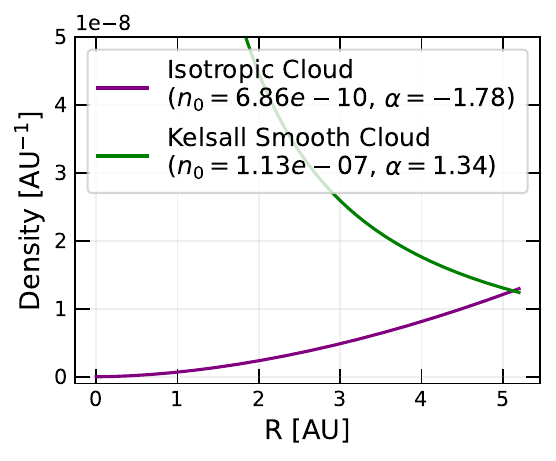}
    \caption{IPD density versus distance from the Sun ($R$) for a proposed isotropic component, compared with that from the smooth cloud component of the Kelsall model along the ecliptic.}
    \label{fig:density_vs_R}
\end{figure}

The MCMC run used 15 walkers for 12,000 iterations. The walkers converged within $\sim$2,000 iterations, and we adopted a burn-in of 4,000 iterations.
The best-fit parameters (with $1\sigma$ uncertainties) are $n_0 = (6.86 \pm 3.25) \times 10^{-10}$ AU$^{-1}$ and $\alpha = -1.78 \pm 0.44$. Figure \ref{fig:density_vs_R} compares the resulting dust density distribution to the original Kelsall smooth cloud. Given the negative power-law index, emission from dust within 1 AU is likely negligible, implying that the isotropic component is effectively shell-like with a mean radius of $\sim$4--5 AU.

\begin{figure}
    \centering
    \includegraphics[width=\linewidth]{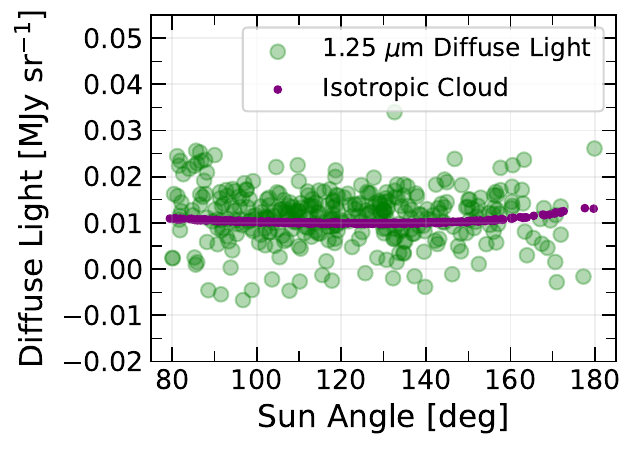}
    \caption{1.25 \micron\ brightness versus Sun Angle for (a) a spherical isotropic-like cloud (purple) with zero uncertainties and (b) current SKYSURF diffuse light measurements (green).}
    \label{fig:iso_cloud_compared_w_f125w}
\end{figure}

Figure \ref{fig:iso_cloud_compared_w_f125w} shows the predicted isotropic component (assuming zero uncertainty) alongside our 1.25 \micron\ diffuse light measurements. The observed brightness from the isotropic component is plausible because the column density varies little with Sun angle, in which case the phase function primarily determines the shape of the profile. Because the large shell radius restricts the sampled scattering angles to a narrow range, the averaged phase function becomes nearly constant, producing an almost isotropic profile aside from the mild Gegenschein enhancement at large Sun angles.

Adding this isotropic component to \obrienzodi\ results in a model that performs extremely well. The filled teal stars in Figure \ref{fig:comparing_to_other_models} shows how our zodiacal light model performs with this isotropic component added. Similarly, the solid green histogram in Figure \ref{fig:comparing_distribution_models_125m_pretty} shows the distribution of the model with the isotropic component added at 1.25 \micron.

These results suggest that a spherical-like component of the IPD could exist and appear isotropic from Earth. However, this remains a toy model designed only to test feasibility. Further work is required to explore which dust distributions are physically plausible and to identify observational signatures that could confirm the presence of a spherical shell. For example, the Fraunhofer absorption method for measuring the absolute intensity of zodiacal light \citep[\eg][]{Korngut_2022} could help isolate an isotropic component. \new{In addition, observations from interplanetary space probes can help verify the existence of such a component, for example, through a higher–spatial-resolution version of the New Horizons mission \citep{Doner_2024} or through new technologies such as solar-sail spacecraft \citep[e.g.,][]{Matsuura_2013}.} Spectroscopic data from SPHEREx or JWST could provide valuable constraints. This isotropic component is included as an optional addition in \obrienzodi. 


\section{Conclusion}

We present an updated zodiacal light model (\obrienzodi) optimized for optical wavelengths (0.3--1.6 \micron). \obrienzodi\ builds upon the \citet{Kelsall_1998} model by modifying the solar irradiance spectrum, albedo, and scattering phase function to match HST sky-SB measurements, while still matching the Kelsall model at 1.25 \micron. \obrienzodi\ is made publicly available on the \href{https://github.com/rosaliaobrien/skysurf/}{SKYSURF GitHub repository}.

To create the model, we use sky-SB measurements from the SKYSURF project, which provide over 5,000 reliable HST measurements across 0.2–1.6 \micron\ with carefully quantified uncertainties and a robust flagging system to ensure only the most reliable HST measurements are used. The final quality-controlled sky-SB sample span a wide range of scattering angles ($80\degree < \epsilon < 180\degree$) and form the basis for optimizing the scattering phase function of the model.

To properly subtract DGL from sky-SB measurements, we present a new DGL estimator based on the 350 and 550 \micron\ dust maps, building on the zodiacal-light–free measurements of \citet{Postman_2024}. Our method explicitly accounts for variations in scattering properties with optical depth. Sky-SB measurements used to create \obrienzodi\ are restricted to those with low DGL levels ($\lesssim0.003$ \MJy).

The most significant improvement to the Kelsall model is our explicit constraint of the wavelength-dependent albedo and scattering phase function, the primary factors determining optical zodiacal light intensity. Our updated zodiacal light model reproduces HST sky-SB data with flat residuals across Sun angle and ecliptic latitude, suggesting that the scattering physics and dust distribution included in the model are sufficient. Extra dispersion at low ecliptic latitudes suggests a possible missing component here.

While other models either overpredict at optical wavelengths or lack spatial/spectral flexibility, \obrienzodi\ delivers consistently reliable performance.  As shown in Figure \ref{fig:comparing_to_other_models}, this yields the most accurate optical zodiacal light model to date, with strong improvements at $\lambda < 1$ \micron. The absolute uncertainties of \obrienzodi\ are $\sim$4.5\%, with an average standard deviation of 0.006 \MJy\ for the residuals (HST Sky -- DGL -- \obrienzodi).

After subtracting our model from sky-SB measurements (HST Sky -- DGL -- \obrienzodi), we detect diffuse light of unknown origin: $0.013 \pm 0.006$ \MJy. This measurement is about twice the predicted EBL. We argue that it may be due to a very dim missing spherical shell-like component to the IPD cloud: a component that appears nearly isotropic from Earth.

We put our measurements of the albedo and scattering phase function in context with the composition of IPD grains. Our measurements suggest that the IPD cloud is composed primarily of low-albedo, small grains, with a best-fit geometric albedo of $p_V \sim 0.07$ that points toward a dominant contribution from comets. In future work, finer spectral resolution can resolve the exact wavelength dependence of the albedo and phase function, and unlock more clues about the composition of IPD.


\obrienzodi\ enables more accurate zodiacal light subtraction for current and future missions. This model directly supports foreground removal and observation planning for large-area sky surveys, including the Euclid space telescope, the Roman Space Telescope, the SPHEREx mission, and the proposed Habitable Worlds Observatory (where understanding exozodiacal phase functions is key for interpreting exoplanet atmospheres). Many of these missions currently rely on Kelsall- or Wright-based models (Figure \ref{fig:comparing_to_other_models}); adopting \obrienzodi\ can substantially improve telescope efficiency, potentially halving required exposure times. It also benefits ongoing missions such as HST and JWST, which have faced challenges from inaccurate foreground predictions \citep[e.g.,][]{Rigby_2023}.

This work represents a major step toward building a comprehensive and reliable model of the diffuse sky. Looking ahead, broader sky coverage (Euclid, SPHEREx, Roman) will help constrain the spectral dependence of dust albedo and scattering phase functions, reveal variations across different components of the IPD cloud, and enhance DGL modeling at optical wavelengths. Spectra from JWST and SPHEREx can also constrain the absolute brightness of zodiacal light (e.g., through Fraunhofer line depths) and may even identify polycyclic aromatic hydrocarbons in the cloud for the first time. Finally, zodiacal light has been shown to be polarized \citep[\eg][]{Takimoto_2022, Takimoto_2023}, and any additional spherically symetric components to the IPD cloud are possibly detectable with polarimetry \citep[][]{Bryden_2023}. The JWST Cycle 3 program SKYSURF-IR will extend the present work using over 65,000 public NIRCam and NIRISS images through JWST Archival Legacy Calibration project project ``SKYSURF-IR'' (R. Ortiz, 2025, in preparation).

\begin{acknowledgements}

\new{The data presented in this article were obtained from the Mikulski Archive for Space Telescopes (MAST) at the Space Telescope Science Institute. The IPPPSSOOT ID lists of the specific observations downloaded for this project are available on MAST via \dataset[doi: 10.17909/ajha-4009]{https://doi.org/10.17909/ajha-4009}. In addition, the sky-SB measurements corresponding to each image ID can be found on our official SKYSURF website: \url{skysurf.asu.edu}.}

Support for HST program 15810 was provided by NASA through a grant from the Space
Telescope Science Institute (STScI), which is operated by the Association of Universities for Research in Astronomy, Inc., under NASA contract NAS 5-26555. These data are associated with HST Archival Legacy Calibration program 15810, and we thank STScI for SKYSURF grants HST-AR-15810. We also thank the STScI MAST Archive staff for their diligent help in getting many Terabytes of SKYSURF data over to ASU, and get our products ported back to MAST. Support for the JWST program 4695 was provided by NASA through a grant from STScI.

RAW, SHC and RAJ also acknowledge support from NASA JWST Interdisciplinary Scientist grants NAG5-12460, NNX14AN10G and 80NSSC18K0200 from GSFC. Work by RGA was supported by NASA under award numbers 80GSFC21M0002 and 80GSFC24M0006.

We acknowledge the indigenous peoples of Arizona, including the Akimel O'odham (Pima) and Pee Posh (Maricopa) Indian Communities, whose care and keeping of the land has enabled us to be at ASU's Tempe campus in the Salt River Valley, where much of our work was conducted.

\new{We thank the anonymous ApJ reviewer for their helpful comments, as these comments helped improve clarity and resulted in a more robust paper. We thank Dr. Shuji Matsuura for the helpful discussion regarding previous work in zodiacal light and diffuse light analysis.}
We thank the SPHEREx Team for hosting RO at Caltech and discussing this work, and the participants of the Aspen Diffuse Backgrounds Conference (March 2024) for valuable conversations. We are grateful to Alex van Engelen and Yogesh Mehta for comments on CIB separation and DGL in GNILC maps, Hypatia Meraviglia for guidance on albedo references, and Sanchayeeta Borthakur and Allison Noble for insightful discussions during RO's defense.

\end{acknowledgements}

\appendix

\section{DGL Improvements} \label{app:dgl_improvements}

Many studies rely on zodiacal light subtraction to isolate and study the DGL. By improving our understanding of zodiacal light at optical wavelengths, this work can contribute to refining DGL models and enhancing the interpretation of full-sky observations.

\begin{figure}[h]
    \centering
    \includegraphics[width=0.7\linewidth]{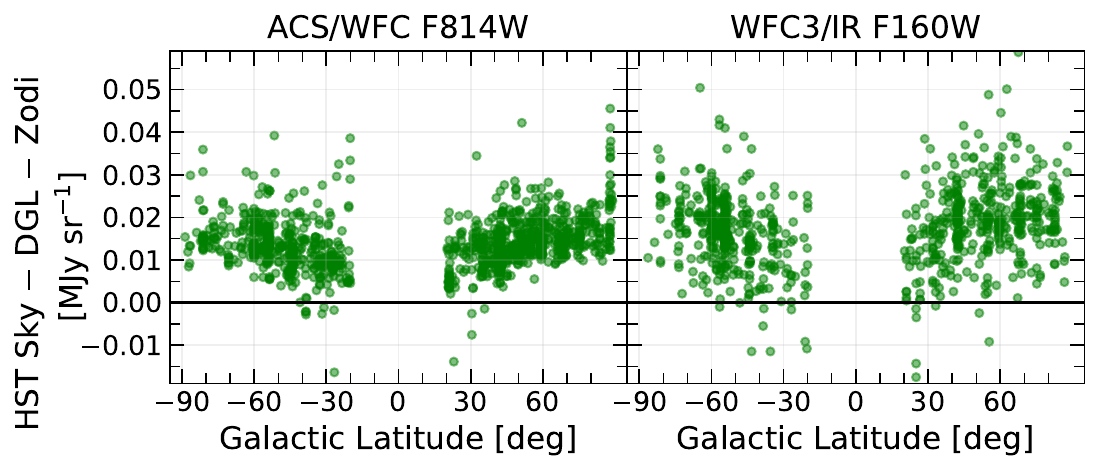}
    \caption{Same as Figure \ref{fig:resid_vs_sunang}, but versus Galactic latitude. On the left, we show the ACS/WFC F814W filter, and on the right, we show the WFC3/IR F160W filter.}
    \label{fig:resid_vs_gallat_pretty}
\end{figure}

Importantly, residual trends in diffuse light as a function of Galactic latitude remain, where our DGL model relatively oversubtracts closer to the Galactic plane, as shown in Figure \ref{fig:resid_vs_gallat_pretty}. It is likely that the function \( g(\gallat) \) (Equation \ref{eq:final_dgl_asym}) or the scattering asymmetry factor ($f_{\lambda}$) assumed in this work need to be improved.
Moreover, \citet{Sano_2025} highlight that the DGL at optical wavelengths is not yet fully understood in regions with denser column densities, noting that the DGL intensity likely decreases with increasing optical depth due to self-absorption. This is not considered in this work, but represents a valuable avenue for future studies. Large surveys from telescopes such as Roman, Euclid, and SPHEREx will help advance this effort by mapping DGL at optical to near-infrared wavelengths deeper than done before. 

\section{Public Code}\label{app:public_code}

The model is distributed as \new{both Python and IDL packages on the \href{https://github.com/rosaliaobrien/skysurf/}{SKYSURF GitHub repository}}. \obrienzodi\ can generate zodiacal light predictions at any day of the year, sky position, and wavelength from 0.25 to 1.7 \micron. Functionally, the package is a copy of the original Kelsall model IDL routine, modified only by the changes described in this paper. The final albedo and phase function lines follow Equations \ref{eq:final_albedo}–\ref{eq:final_w3}. The public code has an option to run it following the original Kelsall formulation.

Importantly, the public code should only be considered reliable (following the uncertainties in Section \ref{sec:uncertainties}) between wavelengths 0.3--1.6 \micron\ and for Sun angles greater than $80\degree$. Outside that wavelength window the code still runs, but it holds the albedo and phase function parameters fixed to their boundary values at 0.25 \micron\ or 1.7 \micron, following Equations \ref{eq:final_albedo}-\ref{eq:final_w3}. Because this work treats only the scattering component of the IPD cloud, it is most reliable at wavelengths $\lesssim$3 \micron; beyond that, thermal emission dominates and users should revert to the original Kelsall model or the Wright model.


Solar irradiance values are provided between 0.2 and 4.9 \micron\ in 0.05 \micron\ increments, where the TSIS spectrum is folded with a simple box bandpass with a width of 0.1 \micron. Since the TSIS-1 HSRS spectra used in this work only span $202-2730$ nm in wavelength, for wavelengths greater than 2.5 \micron, we use the MODTRAN Extraterrestrial Spectra\footnote{\url{https://www.nrel.gov/grid/solar-resource/spectra}} from the National Renewable Energy Laboratory. Specifically, we use their Cebula+Kurucz dataset. We compare filters with similar central wavelengths but different bandpasses (e.g. ACS/WFC versus WFC3/UVIS) and find that built-in irradiance spectrum is accurate to $\sim$2\%. If a project demands more accurate predictions (\ie\ $<$2\% accuracy), users may supply their own irradiance spectrum directly (\eg\ one already folded with a given bandpass) to the routine.


Finally, the original model implements a color correction, which is described in Section 5.5 of the COBE Diffuse Background Experiment (DIRBE) Explanatory Supplement\footnote{\url{https://lambda.gsfc.nasa.gov/data/cobe/dirbe/doc/des_v2_3.pdf}}. The DIRBE data are calibrated assuming the source spectrum $\nu I_{\nu}$ is constant. The original Kelsall model was modified to follow this DIRBE data prescription. Therefore, for our public model, we turn off the color correction. This color correction is only applied to the thermal component of the model ($\lambda \gtrsim 3.5$ \micron), and does not affect HST's wavelengths. This aspect will need to be reconsidered when expanding these models over the full JWST NIRcam and NIRISS wavelength range of 0.7-5 \mum\ for SKYSURF-IR.


\section{Comparing with Geometric Albedos} \label{app:geometric_albedo}


There are a few different definitions of albedo, as reviewed in \cite{Hanner_1981}. Differing definitions may lead to misleading interpretations of composition properties, due to things like shadow-hiding opposition effects (there are smaller shadows on individual grains at lower phase angles, resulting in a brighter surface) and assumptions about the scattering phase function \citep[\eg][]{Beck_2021}.

The most general definition of albedo is that $A=Q_{\text{sca}}/Q_{\text{ext}}$, where $Q_{\text{sca}}$ is the scattering efficiency and $Q_{\text{ext}}$ is the extinction (scattering$+$absorption) efficiency \citep[\eg\ Eq. 22.1 of][]{Draine_2011_book}. The ``Bond albedo'' is defined to be the total reflected light across the entire electromagnetic spectrum. Finally, the ``geometric albedo'' ($p_V$, with the $V$ corresponding to visual wavelengths) is the ratio of the reflected light to that of a Lambertian disk, always observed at a phase angle of 0$\degree$. Asteroid studies in particular often use the geometric albedo \citep{Lebofsky_1986}, while the Kelsall model albedo is closer to the general definition of albedo for individual particles.

We follow \cite{Lumme_1985} to relate the single-scattering albedo to geometric albedo.  They relate $A$ to $p_V$ by

\begin{equation}
    p_V = \frac{A}{4}\Phi_L(\pi),
\end{equation}

\noindent where $\Phi_L$ is their definition of the phase function, which is normalized differently then ours and is different by a factor of 4$\pi$. Therefore, we related relate $A$ to $p_V$ by

\begin{equation}
    p_V = A\pi\Phi(\pi).
\end{equation}

\section{HST vs COBE} \label{app:comparing_hst_and_cobe}

COBE provided nearly year-round, full-sky coverage with its DIRBE instrument, which operated at very low temperatures for ten months, mapping the entire sky multiple times in the first six months and covering most of it again in the following four. In contrast, although HST observes only a small portion of the sky, it spans a wider range of Sun angles, allowing it to constrain the scattering properties of the IPD cloud independently.

\begin{figure*}
    \centering
    \includegraphics[width=\linewidth]{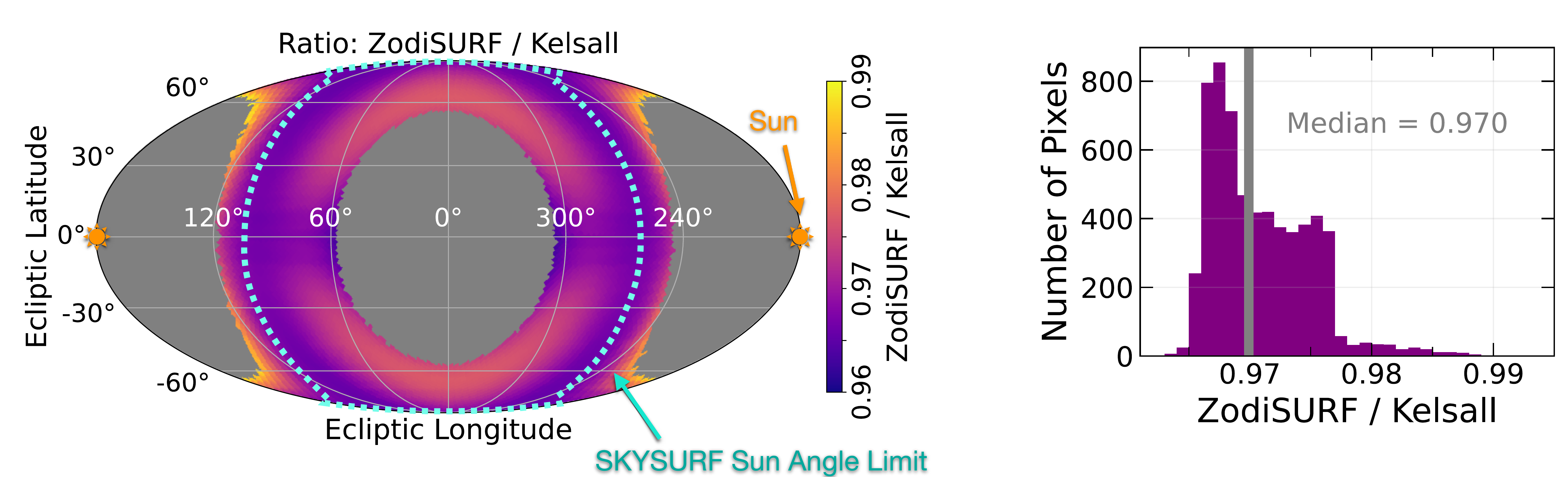}
    \caption{2D representation of the ratio of \obrienzodi\ over the Kelsall model at 1.25 \micron. We mask HEALpix pixels (nside$=32$) outside of COBE's Sun Angle range, as COBE was limited to Sun Angles between 64\degree\ (outer annulus) and 124 \degree (inner annulus). We show the SKYSURF Sun angle limit for comparison.  For SKYSURF, we only use images taken with Sun angles $>80\deg$, which represents all sky within of the cyan dashed annulus. On the right panel, we show a histogram of all unmasked pixels, showing that the median ratio between the two models is 0.970. To make this figure, we fixed the day number to be 267 so that the Sun sits at an Ecliptic Longitude $\sim$180\degree. }
    \label{fig:2d_ratio_comparison}
\end{figure*}
\begin{figure*}
    \centering
    \includegraphics[width=\linewidth]{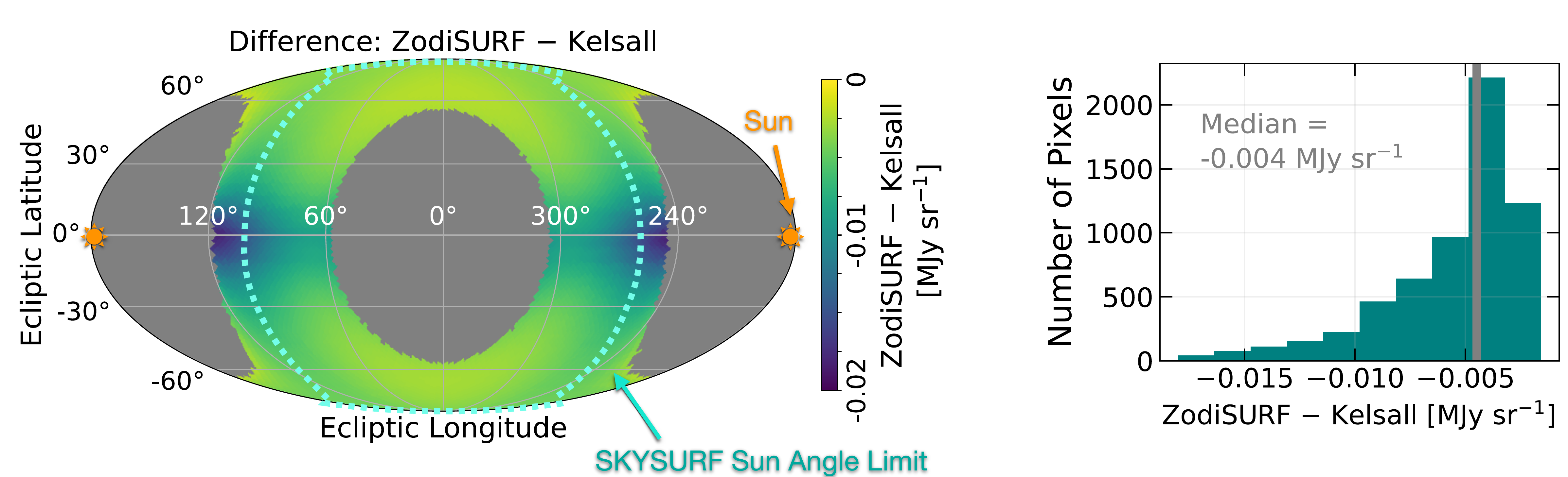}
    \caption{Same as Figure \ref{fig:2d_ratio_comparison}, except showing the difference between \obrienzodi\ and the \citet[][]{Kelsall_1998} model. On the right panel, we show a histogram of all unmasked pixels, showing that the median difference between the two models is 0.004 \MJy.}
    \label{fig:2d_diff_comparison}
\end{figure*}

Figures \ref{fig:2d_ratio_comparison} and \ref{fig:2d_diff_comparison} show the ratio and difference between the two models, fine-tuned at COBE's Sun angle coverage. Specifically, Figure \ref{fig:2d_ratio_comparison} demonstrates that both models agree within 4\% in the appropriate Sun angle range. The structure seen in Figure \ref{fig:2d_ratio_comparison} corresponds to differences in the phase function shown in Figure \ref{fig:phasefuncTalbedo_colored_wave_wkelsall}. Figure \ref{fig:2d_diff_comparison} highlights how the relative changes in the models can result in noticeable differences in the absolute zodiacal light brightness near the ecliptic plane, because zodiacal light is much brighter there.

Table 7 of \cite{Kelsall_1998} quotes an uncertainty at 1.25 \micron\ to be 15 \nW, which corresponds to 0.006 \MJy. This uncertainty is measured by comparing different cloud models, such as a simple ellipsoid, a sombrero model \citep{Giese_1986}, and the widened fan model they determined to be the best fit. They took the two models with the largest difference in the region of the sky with the largest variation, and quantified that variation into the uncertainty they list in Table 7. For comparison, the median difference between \obrienzodi\ and the \citet[][]{Kelsall_1998} model is -0.004 \MJy (Figure \ref{fig:2d_diff_comparison}).



\bibliography{ref}{}
\bibliographystyle{aasjournal}

\end{document}